\documentclass[aps,prb,reprint,longbibliography,superscriptaddress]{revtex4-2}
\usepackage{mathrsfs}
\usepackage{amsmath,gensymb}
\usepackage{amsfonts}
\usepackage{amssymb}
\usepackage{amsthm}
\usepackage{graphicx}
\usepackage{natbib}
\usepackage{color}
\usepackage{hyperref}
\usepackage{bm}
\usepackage[caption=false]{subfig}
\usepackage{verbatim}
\usepackage[normalem]{ulem}
\usepackage{soul}
\usepackage[table]{xcolor}
\usepackage{multirow}

\begin{document}
\title{A unified tight-binding description of the electronic structure and Ising protection of superconductivity in misfit
layered compounds}

\author{G. A. Bobkov}
\email{gabobkov@mail.ru}
\affiliation{Moscow Institute of Physics and Technology, Dolgoprudny, 141700 Moscow region, Russia}

\author{I. A. Shvets}
\affiliation{Laboratory of Nanostructured Surfaces and Coatings, Tomsk State University, 634050 Tomsk, Russia}

\affiliation{Moscow Institute of Physics and Technology, Dolgoprudny, 141700 Moscow region, Russia}

\author{I.V. Bobkova}
\email{ivbobkova@mail.ru}
\affiliation{Moscow Institute of Physics and Technology, Dolgoprudny, 141700 Moscow region, Russia}

\begin{abstract}
Misfit layered compounds (MLCs) offer a unique bulk platform for realizing exotic quantum states typically associated with two-dimensional transition-metal dichalcogenides (TMDs), most notably Ising-protected superconductivity. Yet a theoretical description capturing their electronic structure beyond the simplistic picture of electronically isolated TMD layers has been lacking. Here, we develop a unified tight-binding model for metal dichalcogenide-based MLCs, parameterized by extensive density-functional theory (DFT) calculations across multiple structural configurations and chemical compositions. We show that the intervening tetragonal layers play an active role beyond charge reservoirs: they mediate a significant interlayer spin-orbit coupling entirely absent in the standard rigid-band picture. This emergent interlayer spin-orbit coupling is essential for reproducing the DFT band structure of bulk MLCs and, when incorporated into Bogoliubov--de Gennes calculations, provides a microscopic mechanism for the Ising protection of superconductivity by strongly enhancing the in-plane critical field. Our framework establishes MLCs as a distinct class of three-dimensional materials with intrinsically coupled layers and
emergent spin-orbit phenomena.

\end{abstract}

\maketitle

\section{Introduction}

Misfit layered compounds (MLCs) are a class of heterostructured materials comprising two interleaved, chemically distinct substructures that form a commensurate superstructure along one in-plane direction while remaining incommensurate along another~\cite{Ng2022,Aliev2020}. Typically, one subsystem adopts a cubic rock-salt-type structure (T-layers), while the other forms a hexagonal or trigonal layered framework (H-layers). To date, the most extensively studied MLCs are based on either layered cobaltites~\cite{Boullay1996,Miyazaki2000,Karppinen2004,Panchakarla2016} or metal dichalcogenides~\cite{Wiegers1996,Aliev2020,Ng2022,Khadiev2024}. A remarkable variety of correlated electronic phenomena, including complex magnetism and superconductivity, have already been realized in these systems. From a structural perspective, MLCs represent a unique class of materials that naturally realize van der Waals heterostructures, offering an intrinsic and scalable route to engineer electronic properties at the atomic-layer level.

In the context of superconductivity, metal dichalcogenide-based MLCs have attracted intense interest in recent years \cite{Samuely2021,Leriche2021,Samuely2023,Luo2016,Wiegers1992,Lafond1997,Nader1998,Jiang2024,Samuely2025,Wang2025,Zullo2024,Sajilesh2025}. These compounds are composed of alternating layers described by the general formula [(MX)$_{1+\delta}$]$_{m}$[TX$_{2}$]$_{n}$ (where M = Sn, Pb, Bi, Sb, Y, rare earths; T = Sn, Nb, Ta, Ti, V, Cr; X = S, Se, Te; and $m$, $n$ are integers)~\cite{Rouxel1995,Wiegers1996,Ng2022,Aliev2020,Khadiev2024}. Here, the misfit parameter, $\delta$, describes the ratio between the lattice parameters along the incommensurate crystallographic $a$-axis. The primary interest in these materials stems from the fact that, despite being bulk three-dimensional crystals, they can exhibit the unique electronic properties characteristic of individual monolayers of transition-metal dichalcogenides (TMDs).

Owing to the broken inversion symmetry of 1H-TMD monolayers, they possess Ising-type spin-orbit coupling (ISOC). In such monolayers, the ISOC locks the electron spins to a direction strictly out-of-plane, an effect that can be conceptualized as an enormous effective pseudo-magnetic field~\cite{Lu2015,Xi2016,Saito2016}. This ISOC is responsible for the phenomenon of Ising superconductivity \cite{Ovchinnikov2025}, whereby singlet Cooper pairs are protected from pair-breaking in an in-plane external magnetic field. The out-of-plane spin polarization prevents the Zeeman-driven spin flip that would otherwise destroy the singlet pair \cite{Sarma1963,Maki1964}, enabling the superconducting state to persist to in-plane critical fields $H_{c,\parallel}$ that profoundly exceed the Pauli paramagnetic limit~\cite{Clogston1962,Chandrasekhar1962}. Experimentally, the Ising protection of superconductivity has been observed to weaken with increasing film thickness, yet it survives even in films with an even number of monolayers---despite the formal recovery of inversion symmetry in such systems \cite{Xi2016,delaBarrera2018,Simon2024}.

Potentially more far-reaching than the enhanced resilience against magnetic fields is the prospect of harnessing ISOC as a design principle for realizing a broad spectrum of exotic quantum states. These include equal-spin Andreev reflections~\cite{Zhou2016}, topological superconductivity~\cite{Hsu2017,He2018,Glodzik_2020,Shaffer2020}, Majorana bound states~\cite{Zhou2016,Sharma2016,Zhang2016}, the layer-selective Fulde-Ferrell-Larkin-Ovchinnikov state emerging at high in-plane fields~\cite{Itahashi2025}, and non-unitary pairing states capable of sustaining dissipationless spin currents~\cite{Bobkov2024_spin}.

Albeit a fascinating phenomenon, the long-standing assumption that ISOC is fundamentally restricted to two-dimensional and quasi-two-dimensional structures has presented a significant challenge for both in-depth experimental investigation and practical applications. Probing the intrinsic physics of mechanically exfoliated monolayers is often hindered by sample degradation, limited dimensions, and low total responses to different probes---for instance, specific heat, nuclear magnetic resonance spectroscopy, and inelastic neutron scattering are all effectively inapplicable to single monolayers due to their vanishingly small scattering volumes. In this regard, MLCs that intrinsically host Ising-superconducting building blocks offer a transformative advantage: a bulk, three-dimensional crystalline platform that is generally more stable, homogeneous, and amenable to a wider array of bulk-sensitive experimental probes, while preserving the protected physics of the two-dimensional limit.

In the existing literature on MLCs, this remarkable behavior is often attributed to the spatial separation of the TX$_2$ slabs by the intervening MX layers, which is assumed to electronically isolate each TX$_2$ layer, allowing it to behave as an effective monolayer embedded within a three-dimensional bulk matrix \cite{Samuely2021,Samuely2023,Zullo2023,Leriche2021,Zullo2024,Sajilesh2025}. However, as we will demonstrate below, this seemingly natural assumption does not fully capture the underlying physics. Building on the electronic properties of free-standing monolayers and bilayers, we develop a comprehensive model of MLCs that accurately captures the electronic structure in the vicinity of the Fermi surface for arbitrary stoichiometries [(MX)$_{1+\delta}$]$_{m}$[TX$_{2}$]$_{n}$, and is equally applicable to mixed compositions of the form $\rm M \to (\rm M_1)_z (\rm M_2)_{1-z}$. A central physical insight emerging from our model is that the tetragonal layers exhibit a dual effect: they partially suppress the interlayer coupling between adjacent trigonal blocks and generate a remarkably strong interlayer spin-orbit coupling.

This finding lies outside the scope of the conventional ``rigid-band'' picture, which treats an MLC as an assembly of electronically isolated trigonal layers, with the sole role of the tetragonal layers being a shift of the chemical potential in the trigonal subsystem via interfacial charge transfer~\cite{Zullo2023}. In particular, our model naturally accounts for the Ising protection of superconductivity observed in (LaSe)$_{1.14}$(NbSe$_2$)$_2$ \cite{Samuely2021,Samuely2023}, where, according to the isolated-bilayer scenario, spin splitting should be entirely absent due to the presence of an inversion center within each NbSe$_2$ bilayer.

The remainder of the paper is organized as follows. Section~\ref{sec:symmetry_analysis} presents a general symmetry analysis of the possible types of interlayer coupling between TMD monolayers. In Sec.~\ref{sec:free_standing}, we summarize the electronic properties of isolated TMD bilayers with different stacking sequences, focusing on NbSe$_2$, and introduce a minimal model Hamiltonian that captures their essential features. Section~\ref{sec:1T1H} presents a detailed DFT analysis of the electronic structure of several 1H1T ($m=n=1$) MLCs based on NbSe$_2$, from which we construct a model Hamiltonian in momentum space for this class of materials. In Sec.~\ref{sec:1T2H}, the model is generalized to more complex MLC structures. In Sec.~\ref{sec:tight-binding}, we recast all previously introduced Hamiltonians in a tight-binding form, thereby obtaining a unified model that is valid throughout the entire Brillouin zone and is readily applicable to spatially inhomogeneous problems. Then the model is validated against DFT calculations. Finally, in Sec.~\ref{sec:ising_protection}, we show that the interlayer spin-orbit coupling inherent to our model naturally accounts for the experimentally observed Ising protection of superconductivity in 1H1T and 2H1T MLCs. 

\section{Symmetry analysis of allowed interlayer couplings between trigonal
layers}
\label{sec:symmetry_analysis}

Our goal is to develop a universal description of the electronic structure near the Fermi surface for a broad class of MLCs. Throughout this work, we restrict our analysis to the bands stemming from the Fermi-level states of an isolated TMD monolayer, as these states govern the low-energy physics.

\begin{figure}[!tbh]
\includegraphics[width=\columnwidth]{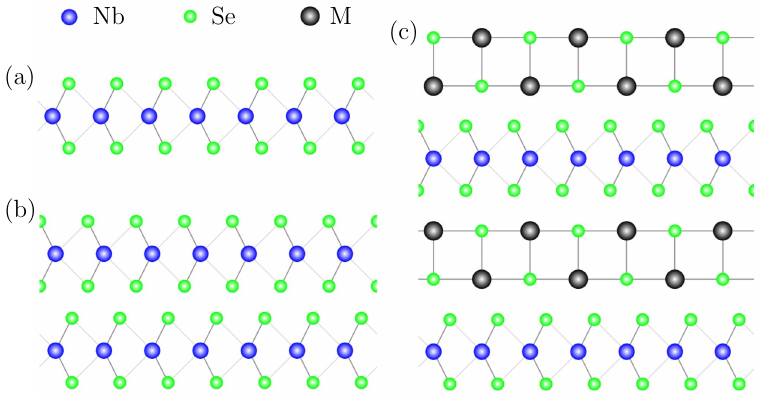}
\caption{Atomic structure of (a) 1H-NbSe$_2$ monolayer and (b) 2H-NbSe$_2$ bilayer (side view).  (c) Representative example of a structure where two NbSe$_2$ monolayers are coupled via an intervening tetragonal (rock-salt) layer, as realized in misfit layered compounds.} 
 \label{fig:atomic_structure}
\end{figure}

Achieving this goal requires two essential ingredients: (i)~the electronic structure of an isolated TMD monolayer and (ii)~the nature of its coupling to adjacent TMD monolayers and to the intervening tetragonal layers in various heterostructures and stacking configurations. The atomic structure of a 1H-NbSe$_2$ monolayer is shown in Fig.~\ref{fig:atomic_structure}(a), together with two representative coupling scenarios: NbSe$_2$ monolayers coupled directly to each other [Fig.~\ref{fig:atomic_structure}(b)] and coupled via an intervening tetragonal layer [Fig.~\ref{fig:atomic_structure}(c)]. In this section, we examine the symmetry-allowed types of interlayer coupling between TMD monolayers.

In the absence of intrinsic magnetism, external magnetic field, and Rashba-type spin-orbit coupling, the spin quantization axis is set by the direction of the pseudo-magnetic field of the ISOC---that is, the normal to the layer plane, along the $z$~axis. The projection of the electron spin onto this axis is a good quantum number, and therefore, in momentum space, the Hamiltonian of a 1H-TMD monolayer can be written separately for spin-up and spin-down electrons ($\sigma=\pm 1$) as \cite{Yuan2014,Lu2015,Xi2016,Saito2016}
\begin{align}
    \hat H_{\rm 1H}^\sigma = \zeta(\bm p)+\alpha(\bm p)\sigma,
    \label{H1}
\end{align}
where the electron dispersion in the absence of spin–orbit coupling is denoted as $\zeta(\bm p) = \zeta(-\bm p)$, while the intralayer ISOC gives rise to an odd in momentum Ising-type splitting for spin-up and spin-down electrons,  denoted as $\alpha (\bm p) = -\alpha(- \bm p)$. 

The Hamiltonian of any pair of TMD layers, including the interlayer coupling between them (which may occur either directly or via an intervening tetragonal layer), can be written as
\begin{align}
    \hat H^{\sigma}_{\rm 2}=\left(\begin{array}{cc}
\zeta(\bm p)+\sigma\alpha(\bm p) &  V^\sigma \\
V^{\sigma*} &  \zeta(\bm p)\pm\sigma\alpha(\bm p) 
\end{array}\right). 
\label{H2}
\end{align}
This Hamiltonian acts on the two component wave function $(\Psi_{1}^\sigma(\bm p), \Psi_{2}^\sigma(\bm p))^T$ describing spin-$\sigma$ electrons in the interacting layers $1,2$. The sign of the bottom right matrix element in $\hat{H}_2^\sigma$ depends on the relative orientation of the two coupled monolayers. For the 2H-TMD structure [see Fig.~\ref{fig:atomic_structure}(b)], which possesses an inversion center, the upper left and bottom right matrix elements have opposite signs. By contrast, when both layers share the same orientation---corresponding to AA-stacking, i.e., a non-centrosymmetric configuration---the two matrix elements carry the same sign.

Term $V$ accounts for the interlayer interaction. Introducing Pauli matrices $\rho_i$, $i=x,y,z$  in the $(1,2)$-layer space with $\rho_0$ being the unity matrix, the general expression for 
\begin{align}
    \hat V^\sigma =\left(\begin{array}{cc}
0 &  V^\sigma \\
V^{\sigma*} &  0
\end{array}\right)
\label{V_matrix}
\end{align}
can be expanded as
\begin{align}
    \hat V^\sigma=\sum \limits_{i=x,y} [x_{i+}^0(\bm p) + \sigma x_{i+}^\sigma(\bm p) + x_{i-}^0(\bm p) + \sigma x_{i-}^\sigma(\bm p)]\rho_i , 
    \label{V_general}
\end{align}
where $x_{i+}^{0,\sigma}(\bm p)$ and $x_{i-}^{0,\sigma}(\bm p)$ are real functions, which are even and odd in momentum, respectively. The behavior of each term in this expression under time-reversal symmetry ($\mathcal{T}$) is summarized in the first row of Tab.~\ref{tab:symmetry}. For brevity, we introduce the notation $e_p (\bm p) = -e_p (-\bm p)$ for the momentum-odd terms $x_{i-}^{0,\sigma}$. All misfit compounds and TMDs considered in this work are nonmagnetic; their Hamiltonians must therefore be invariant under time reversal. Consequently, the general form of the interlayer coupling $\hat{V}^\sigma$ for such structures can only contain terms marked by a $+$ sign in the first row of Tab.~\ref{tab:symmetry}. With no further restrictions beyond time-reversal symmetry, $\hat{V}^\sigma$ can be written as
\begin{align}
    V^\sigma=v(\bm p)+\nu(\bm p)\sigma-i\lambda(\bm p)\sigma-iw(\bm p),
\label{V_tr}
\end{align}
where the real functions $v(\bm{p})$ and $\lambda(\bm{p})$ are even in momentum, $v(\bm{p}) = v(-\bm{p})$, $\lambda(\bm{p}) = \lambda(-\bm{p})$, while $\nu(\bm{p})$ and $w(\bm{p})$ are odd, $\nu(\bm{p}) = -\nu(-\bm{p})$, $w(\bm{p}) = -w(-\bm{p})$; see also the bottom row of Tab.~\ref{tab:symmetry}, where all four coefficients are matched, for convenience, with their corresponding layer, spin, and momentum structure. Physically, the parameter $v(\bm p)$ governs the conventional spin-independent interlayer hopping, while $\nu(\bm p)$ characterizes the interlayer ISOC. In the bulk crystal, the $\rho_y$-type interlayer terms yield a momentum dependence $\propto p_z$ along the normal to the layers. Consequently, $\lambda (\bm p)$ describes a distinct spin-orbit interaction, in which the electron spin is locked to the out-of-plane momentum $p_z$. Finally, $w(\bm p)$ accounts for a contribution to the spin-independent interlayer coupling that captures the Fermi-surface asymmetry with respect to the layer plane, should such an asymmetry arise in the system under consideration.

\begin{table}
\begin{center}
\begin{tabular}{|c|c|c|c|c|c|c|c|c|c|c|}
\hline
 & $\rho_x$ & $\rho_x\sigma$ & $\rho_x e_p$ & $\rho_x \sigma e_p$ & $\rho_y$ & $\rho_y\sigma$ & $\rho_y e_p$ & $\rho_y \sigma e_p$ \\
\hline
$\mathcal{T}$ & $+$ & $-$ & $-$ & $+$ & $-$ & $+$ & $+$ & $-$ \\
\hline
$I$ & $+$ & $+$ & $-$ & $-$ & $-$ & $-$ & $+$ & $+$ \\
\hline
$\sigma_h$ & $+$ & $+$ & $+$ & $+$ & $-$ & $-$ & $-$ & $-$ \\
\hline
def & $v$ &   &   & $\nu$ &   & $\lambda$ & $w$ &  \\
\hline
\end{tabular}
\end{center}
\caption{\label{tab:symmetry}Transformation properties of the eight elementary interlayer coupling terms in Eq.~\ref{V_general} under time reversal $\mathcal{T}$, inversion $I = \{1\leftrightarrow 2, \,\bm{p}\to -\bm{p}\}$, and mirror reflection $\sigma_h = \{1\leftrightarrow 2\}$. The last row gives the mapping to the coefficient functions $v$, $\nu$, $\lambda$, $w$ introduced in Eq.~\eqref{V_tr}.}
\end{table} 

In the second and third rows of Tab.~\ref{tab:symmetry}, we list the transformation properties of each term in $\hat{V}^\sigma$ under the inversion operation $I = \{ 1 \leftrightarrow 2, \ \bm{p} \to -\bm{p} \}$ and under the mirror reflection $\sigma_h = \{ 1 \leftrightarrow 2 \}$ across the plane bisecting the two layers. The behavior of each term in the interlayer coupling Hamiltonian under these symmetry operations will be exploited in the next section to construct the Hamiltonians of bilayer TMD structures.

\section{Electronic structure of free-standing  $\mathrm{\textbf{NbSe}_2}$ bilayers}
\label{sec:free_standing}

Our model for the electronic structure of misfit layered compounds is built on two essential ingredients: the electronic structure of an isolated transition-metal dichalcogenide monolayer and the interlayer coupling. In this section, we present a systematic analysis of the interlayer coupling in a TMD bilayer. The hopping amplitudes are computed specifically for NbSe$_2$; the conceptual framework, however, is general and directly transferable to other metallic transition-metal dichalcogenides, with the understanding that the precise momentum dependence of all Hamiltonian matrix elements must be recalibrated for each material against DFT reference calculations.

\begin{figure}[!tbh]
\includegraphics[width=0.6\columnwidth]{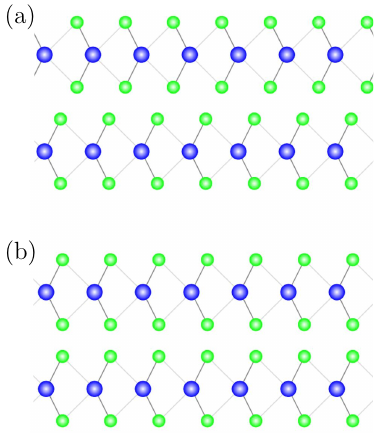}
\caption{(a) Atomic structure of a 2H-stacked TMD bilayer (e.g., NbSe$_2$), which possesses inversion symmetry. (b)~Atomic structure of a hypothetical 1H+1H (AA-stacked) TMD bilayer, which is non-centrosymmetric.} 
 \label{fig:bilayer_structure}
\end{figure}

The atomic structure of the 2H-TMD bilayer is presented in Fig.~\ref{fig:bilayer_structure}(a). Fig.~\ref{fig:bilayer_structure}(b) shows a hypothetical TMD bilayer, which we term as 1H$+$1H-TMD bilayer (also called AA-stacked or non-centrosymmetric \cite{He2014}), consisting of two identical 1H layers stacked without the inversion operation that relates the layers in the 2H polymorph. Although this stacking is not realized in bulk TMDs, its consideration is essential for a comprehensive tight-binding description of misfit layered compounds.

For the bilayer in its 2H-configuration, the Hamiltonian can be expressed as a $2\times 2$ matrix: 
\begin{align}
    \hat H^{\sigma}_{\rm 2H}=\left(\begin{array}{cc}
\zeta(\bm p)+\sigma\alpha(\bm p) &  v(\bm p) \\
v(\bm p) &  \zeta(\bm p)-\sigma\alpha(\bm p) 
\end{array}\right).
\label{ham_2H}
\end{align}
The spin-orbit terms of the $\nu$ and $\lambda$ type are forbidden by inversion symmetry (see Tab.~\ref{tab:symmetry}). The $w$-type term is formally allowed for a 2H-TMD bilayer; nevertheless, we find no physical motivation to introduce it, as the Fermi surface of a 2H-TMD bilayer is symmetric about the layer plane and, as demonstrated below, the DFT electronic spectrum of 2H-NbSe$_2$ is perfectly reproduced by the conventional interlayer hopping $v$ alone.

For the bilayer in the 1H$+$1H configuration, an analogous $2\times 2$ Hamiltonian can be written in a similar manner. In this case, however, the $\lambda$ and $w$ terms are forbidden by the mirror symmetry $\sigma_h$, leaving $v$ and the interlayer ISOC $\nu$ as the only allowed interlayer couplings:
\begin{align}
    \hat{H}^{\sigma}_{\rm 1H+1H} =
    \begin{pmatrix}
        \zeta(\bm{p}) + \sigma\alpha(\bm{p}) & v(\bm{p}) + \sigma\nu(\bm{p}) \\
        v(\bm{p}) + \sigma\nu(\bm{p}) & \zeta(\bm{p}) + \sigma\alpha(\bm{p})
    \end{pmatrix}.
    \label{H_1H1H}
\end{align}
Note that the diagonal elements are symmetric, in contrast to the 2H case. The two models yield the following dispersion relations:
\begin{align}
    \varepsilon_{\rm 2H}^\sigma(\bm{p}) &= \zeta(\bm{p}) \pm
        \sqrt{\alpha(\bm{p})^2 + v(\bm{p})^2},
    \label{e_2H} \\
    \varepsilon_{\rm 1H+1H}^\sigma(\bm{p}) &= \zeta(\bm{p}) +
        \alpha(\bm{p})\sigma \pm |v(\bm{p}) + \nu(\bm{p})\sigma|.
    \label{e_1H+1H}
\end{align}
As expected, the energy spectrum of the 2H bilayer consists of two spin-degenerate branches owing to the absence of a net ISOC: the ISOC has opposite signs on the two constituent monolayers, and the spin splitting therefore cancels out. In contrast, the 1H$+$1H bilayer possesses a nonvanishing net ISOC, which lifts the spin degeneracy.

\begin{figure}[!tbh]
\includegraphics[width=\columnwidth]{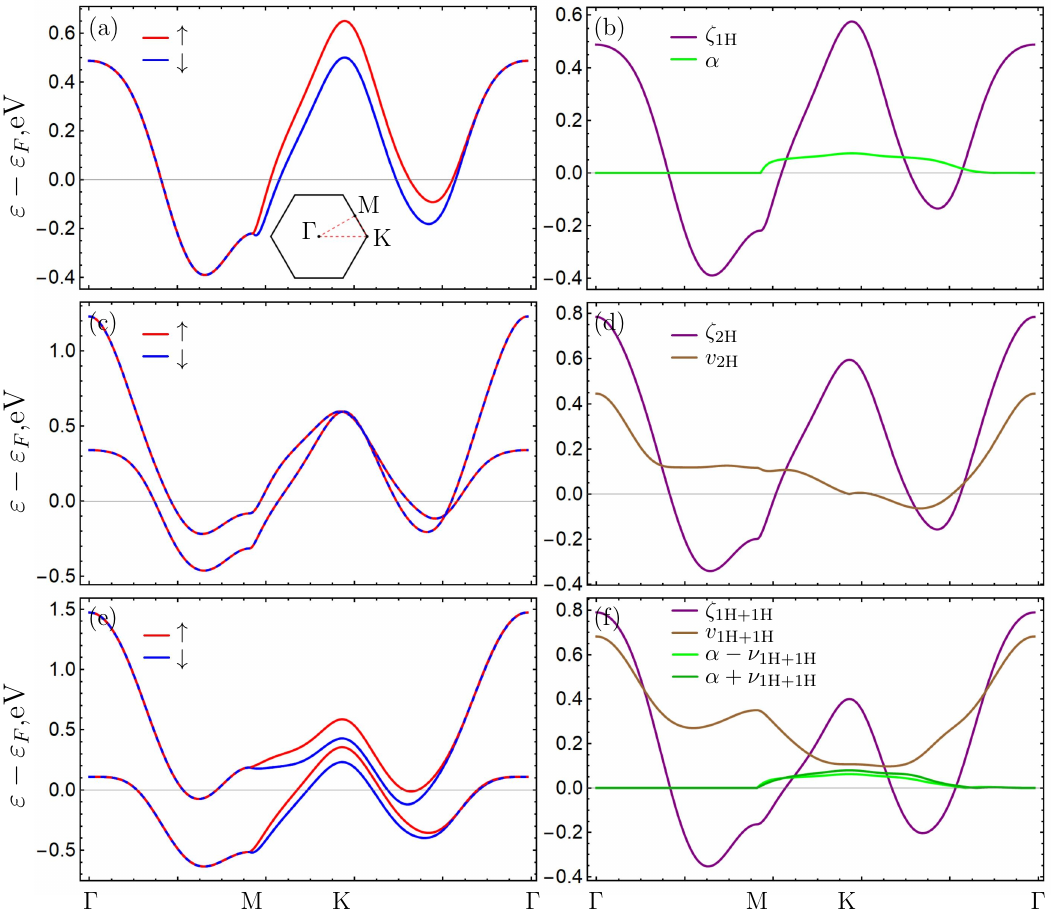}
\caption{DFT-calculated band structures and the corresponding model parameters for the Fermi-surface bands of NbSe$_2$-based systems. (a)~Monolayer 1H-NbSe$_2$: DFT band structure. The inset shows the hexagonal Brillouin zone with the high-symmetry points $\Gamma$, M, and K. (b)~Monolayer parameters $\zeta_{\rm 1H}(\bm{p})$ and $\alpha(\bm{p})$ extracted from the spectrum in (a). (c)~Bilayer 2H-NbSe$_2$: DFT band structure. (d)~Intralayer dispersion $\zeta_{\rm 2H}(\bm{p})$ and interlayer hopping $v_{\rm 2H}(\bm{p})$ extracted from (c). (e)~Bilayer 1H$+$1H (AA-stacked) NbSe$_2$: DFT band structure. (f)~Intralayer dispersion $\zeta_{\rm 1H+1H}(\bm{p})$, interlayer coupling $v_{\rm 1H+1H}(\bm{p})$, and spin-orbit splittings of the upper and lower branches $\alpha(\bm{p}) \pm
\nu_{\rm 1H+1H}(\bm{p})$ extracted from (e).} 
 \label{fig:spectra_free}
\end{figure}

Our next step is to benchmark these minimal models against DFT calculations of the electronic structure of NbSe$_2$ bilayers. The electronic spectra of the 1H-NbSe$_2$ monolayer, the 2H-NbSe$_2$ bilayer, and the 1H$+$1H-NbSe$_2$ bilayer are presented in Figs.~\ref{fig:spectra_free}(a), (c), and (e), respectively. It should be noted that the bands shown originate only from the Nb $d$-orbitals that cross the Fermi level; the full manifold of valence and conduction bands is omitted for clarity. We focus exclusively on these Fermi-surface bands because they govern the low-energy physics of transport phenomena and are directly responsible for superconductivity. 

The DFT calculations were performed using the VASP package \cite{VASP1}. The projector-augmented wave approach \cite{Blochl.prb1994} was used to describe the electron-ion interaction. The generalized gradient approximation (GGA) of Perdew, Burke, and Ernzerhof (PBE) was used as the exchange-correlation functional. The spin-orbit coupling  effect was taken into account. 
Geometry optimization was performed until the residual force on the atoms was less than 20 meV/\AA. In order to describe the van der Waals interactions we made use of the DFT-D3 functional with the Becke-Johnson damping scheme \cite{Grimme2011}.

The DFT spectra presented in Fig.~\ref{fig:spectra_free}(a), (c), and (e) are accurately captured by the proposed minimal models. The parameters $\zeta(\bm{p})$ and $\alpha(\bm{p})$ describing the dispersion of the 1H-NbSe$_2$ monolayer, obtained by fitting the spectrum shown in Fig.~\ref{fig:spectra_free}(a) to Eq.~\eqref{H1}, are displayed in Fig.~\ref{fig:spectra_free}(b). A well-known feature of the electronic spectrum of a metallic nonmagnetic TMD monolayer (NbSe$_2$ being a prototypical example) is the presence of a substantial Ising spin-orbit splitting that reaches maximum amplitude near the K-point. 

The model parameters extracted from the spectra of 2H-NbSe$_2$ and 1H$+$1H-NbSe$_2$ are presented in Figs.~\ref{fig:spectra_free}(d) and (f), respectively. The parameters $\zeta(\bm{p})$ and $\alpha(\bm{p})$ of the bilayer spectra differ only slightly from their monolayer counterparts, confirming that a description rooted in the monolayer electronic structure, supplemented by interlayer couplings, is indeed well justified. As seen from the results, the interlayer Ising-type spin-orbit hopping in 1H+1H-NbSe$_2$ bilayers is very small, although unambiguously present. As we show below, this parameter becomes appreciable in misfit compounds, where its magnitude depends on the presence and chemical composition of the intervening tetragonal layer separating the TMD slabs. A common and characteristic feature observed in all systems studied here (see below) is the momentum-space structure of the interlayer hopping $v(\bm p)$: it reaches its maximum near the $\Gamma$-point and diminishes in the vicinity of the K-point.

Earlier studies put forward a similar minimal model for the 2H-TMD bilayer~\cite{Samuely2023,delaBarrera2018,Engstrom2026}. 
In those treatments, however, the interlayer hopping was approximated as momentum-independent, despite the observation in Ref.~\cite{Samuely2023} that it acquires a momentum dependence across the Brillouin zone. It is also worth noting that an antisymmetric on-site energy shift $\pm \Delta \varepsilon_F$ between the two constituent monolayers was introduced in Refs.~\cite{Samuely2023,Engstrom2026} to account for the breaking of inversion symmetry along the out-of-plane direction induced by the presence of a substrate. This additional term enabled the authors of Ref.~\cite{Engstrom2026} to explain the experimentally observed Ising protection of superconductivity in layered structures with an even number of NbSe$_2$ monolayers \cite{Xi2016,delaBarrera2018,Simon2024}, where the structural inversion symmetry of each 2H-NbSe$_2$ block would otherwise forbid the spin splitting of the spectrum (see Fig.~\ref{fig:spectra_free}(c)).

In the present section, we consider free-standing 2H-NbSe$_2$ bilayers; there is, accordingly, no physical basis for introducing an antisymmetric on-site shift $\pm\Delta \varepsilon_F$. Nor do we find such a justification for the 2H1T bulk MLCs~\cite{Samuely2023}. Nevertheless, the model proposed here is able to account for the experimentally observed enhancement of the critical field in (MX)$_{1+\delta}$[TX$_2$]$_2$ compounds; see Sec.~\ref{sec:ising_protection}.

\section{Electronic structure of 1H1T $\mathrm{\textbf{MLCs}}$: from DFT results to effective model}
\label{sec:1T1H}

We now turn to the construction of a consistent description of the interlayer coupling in misfit layered compounds. Before proceeding to the electronic structure, we briefly summarize the key structural details relevant to our model. We chose MSe (M = Pb, Bi, La) as the tetragonal (T) subsystem. This selection is motivated by the need to cover a broad range of formal valences and charge transfer strengths, while also capturing variations in the structural mismatch parameter. Although both La and Bi possess a formal valence of +3, in contrast to Pb (+2) making it a charge-balanced insulator, the charge transfer from the LaSe subsystem is significantly larger. This is primarily driven by the much lower work function of LaSe compared to BiSe, which facilitates a more pronounced Fermi level upshift in the NbSe$_{2}$ conduction band~\cite{Leriche2021,Zullo2023,Samuely2023}. Additionally, unlike LaSe, the BiSe layers tend to undergo internal structural modulations with the formation of Bi–Bi bonds, which partially traps the excess electron density within the rock-salt slab. Furthermore, compounds based on PbSe, BiSe, and LaSe are well-documented experimentally, providing a solid benchmark for validating our theoretical approach. 

Within a MLC, the NbSe$_2$ layer is subjected to a minor uniaxial tensile strain, which lowers its symmetry from ideal hexagonal to centered orthorhombic, with lattice parameters $a_1 \approx 3.4$\,\AA{} and $b_1 \approx 6$\,\AA{}. The adjacent MSe sublattice is commensurate with the NbSe$_2$ layer along the common $b$-axis ($a_2 \approx b_2 \approx 6$\,\AA{}), but remains incommensurate along the $a$-direction owing to the irrational ratio $|a_1|/|a_2|$. The resulting misfit ratio $a_2/a_1 \approx 7/4$ determines the stoichiometric parameter $\delta \approx 1.14$.

\begin{figure}[!tbh]
\includegraphics[width=\columnwidth]{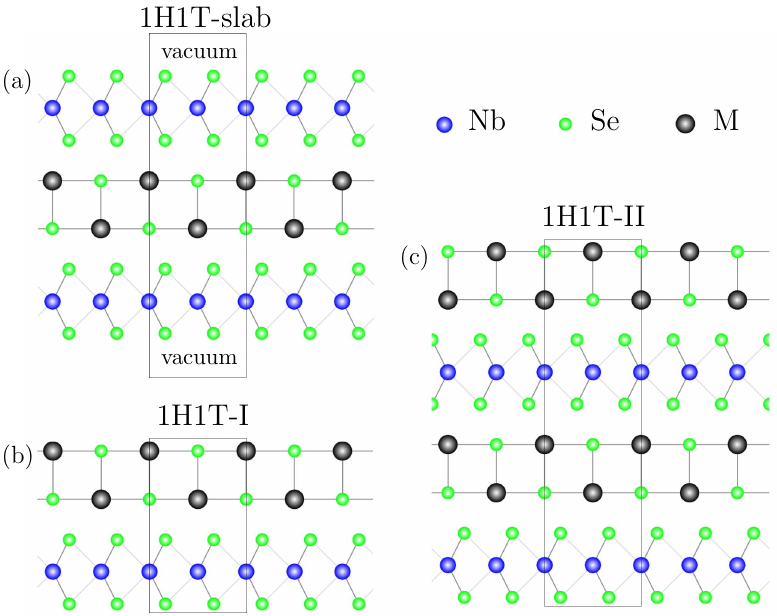}
\caption{Atomic structures (side view, along commensurate \textit{b} direction) of the three MLC configurations
studied in Sec.~\ref{sec:1T1H}: (a)~a misfit slab composed of two trigonal NbSe$_2$ layers separated by an intervening tetragonal MSe layer (1H1T-slab); (b)~a 1H1T bulk MLC (MSe)$_{1.14}$NbSe$_2$ in which all trigonal blocks share the same stacking orientation, so that the unit cell contains a single trigonal layer (1H1T-I); (c)~a 1H1T bulk MLC (MSe)$_{1.14}$NbSe$_2$ in which the unit cell contains two NbSe$_2$ layers arranged in the alternating stacking configuration characteristic of a 2H-NbSe$_2$ bilayer (1H1T-II). The unit cells are indicated by black rectangles.} 
 \label{fig:structures_1H1T}
\end{figure}

We have performed DFT calculations of the electronic structure for three distinct types of MLCs with $m=n=1$, illustrated in Fig.~\ref{fig:structures_1H1T}: (a)~a slab consisting of two trigonal NbSe$_2$ layers separated by an intervening tetragonal MSe layer; (b)~a 1H1T bulk MLC (MSe)$_{1.14}$NbSe$_2$ in which all trigonal blocks share the same stacking orientation, so that the unit cell contains a single trigonal layer (referred to hereafter as 1H1T-I); (c)~a 1H1T bulk MLC (MSe)$_{1.14}$NbSe$_2$ in which the unit cell contains two NbSe$_2$ layers arranged in the alternating stacking configuration characteristic of the 2H-NbSe$_2$ bilayer (referred to hereafter as 1H1T-II). This minimal set of configurations provides the complete information on the electronic spectra needed to unambiguously determine all intralayer and interlayer hopping parameters.

The existence of MLCs of the 1H1T-I and 1H1T-II types has been suggested by experimental reports~\cite{Samuely2021,Samuely2023,Sajilesh2025}. Specifically, a 1H1T-II bulk MLC was reported in Refs.~\cite{Samuely2021,Samuely2023}, while a 1H1T bulk MLC in which all trigonal blocks share the same stacking orientation was reported in Ref.~\cite{Sajilesh2025}. We note that the latter compound, (PbS)$_{1.13}$(TaS$_2$), formally contains two trigonal layers per unit cell owing to a relative lateral shift between adjacent trigonal layers separated by the tetragonal PbS layer. However, our calculations demonstrate that this shift has no discernible effect on the electronic structure; the material is therefore electronically equivalent to a 1H1T-I MLC.

DFT results for the electronic structure of both modifications have been reported for several MLCs~\cite{Samuely2023,Chikina2022}. In this context, it is particularly important to understand how the electronic spectra, and the corresponding effective models derived from them, differ among these configurations. Finally, the slab depicted in Fig.~\ref{fig:structures_1H1T}(a) can be obtained experimentally by exfoliating the corresponding 2H1T misfit compound.

\begin{figure*}[!tbh]
\includegraphics[width=1.8\columnwidth]{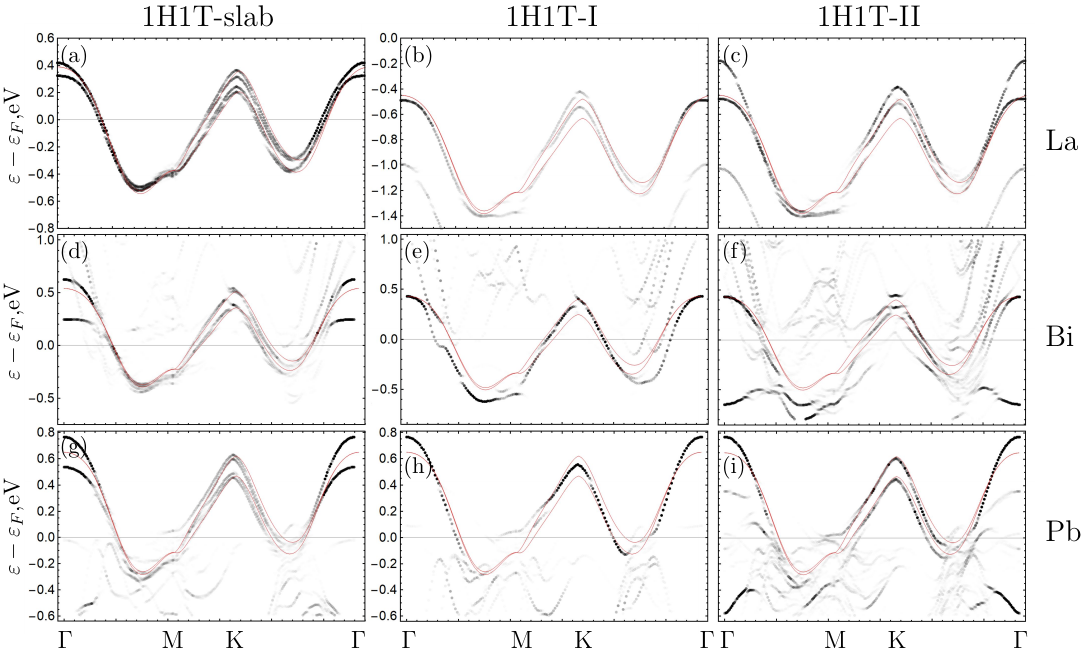}
\caption{Unfolded electronic band structures along the high-symmetry path $\Gamma$--M--K--$\Gamma$ near the Fermi level for 1H1T misfit layered compounds based on NbSe$_2$ and tetragonal MSe layers (M = La, Bi, Pb), calculated within DFT. The columns correspond to the three structural configurations shown in Fig.~\ref{fig:structures_1H1T}: (a,d,g)~1H1T-slab, (b,e,h)~bulk 1H1T-I (all trigonal layers share the same orientation), and (c,f,i)~bulk 1H1T-II (alternating stacking characteristic of the 2H polymorph). The rows denote the chemical composition of the tetragonal layer: La (a--c), Bi (d--f), and Pb (g--i). A thin red line on each panel shows, for comparison, the dispersion of the NbSe$_2$ monolayer subjected to the same uniaxial tensile strain as in the corresponding MLC, but without the surrounding layers. The corresponding bands are rigidly shifted by the charge-transfer-induced energy $\Delta\varepsilon_F$ specific to each structure (see Tab.~\ref{tab:DFT}).} 
 \label{fig:DFT_1H1T}
\end{figure*}

Fig.~\ref{fig:DFT_1H1T} shows the unfolded electronic dispersion near the Fermi level for MLCs  in the three configurations studied here: the 1H1T slab,  1H1T-I, and  1H1T-II  bulk MLCs for M~=~La, Bi, Pb. The unfolding of the band structure of misfit structures was performed using the BandUP code \cite{PRB-Band-UP-2015}. 

A first characteristic feature of these spectra—already noted for certain MLCs (see, e.g., Ref.~\cite{Zullo2023})—is the shift of the Fermi level $\Delta \varepsilon_F$ relative to that of an isolated NbSe$_2$ monolayer, which arises from charge transfer $q$ from the tetragonal layers. The magnitude of the charge transfer (and consequently $\Delta \varepsilon_F$) is directly correlated with the interlayer distance $d_{H-T}$ between NbSe$_2$ and MSe layers (see Table~\ref{tab:DFT}): the smaller the distance, the greater the charge transfer. Notably, in the bulk 1H1T-I and 1H1T-II structures, the charge transfer is twice as large as that observed for the 1H1T-slab configuration. This finding indicates that each tetragonal layer independently donates electrons to the trigonal layer. Among other implications, this suggests that for the surface NbSe$_2$ layer in the bulk MLC, the Fermi-level shift is expected to be identical to that of the 1H1T-slab case. Furthermore, the bulk 1H1T-I and 1H1T-II structures are practically indistinguishable with respect to charge transfer and Fermi-level displacement. 

Another intriguing feature is that a first portion of the transferred charge ($q_0$) does not contribute to the Fermi-level shift but instead participates in the formation of chemical bonds between MSe and NbSe$_2$ layers. Consider the case of M = Pb: at zero Fermi-level shift, there exists a ``baseline'' charge transfer $q_0\sim 0.08$e for the slab, and twice this value for the bulk structures. When Pb is replaced by Bi or La, the entire charge transfer exceeding this baseline gives rise to the Fermi-level shift. Moreover, the shift is governed by the expression 
\begin{align}
    q-q_0 \sim \int\limits_0^{\Delta \varepsilon_F} N(\varepsilon)d \varepsilon,
\end{align}
where $N(\varepsilon)$ is the density of states at energy $\varepsilon$. This implies, in particular, that as long as the charge transfer remains small and $N(\varepsilon)$ may be regarded as constant, the Fermi-level shift is a linear function of the charge transfer with $d(\Delta \varepsilon_F)/dq \sim 1~$eV/e. Conversely, in the bulk 1H1T structure with M~=~La, the charge transfer is so large that the entire conduction band of NbSe$_2$ becomes filled, and the Fermi level moves above it; consequently, $\Delta \varepsilon_F$ increases much more rapidly with increasing $q$.

\begin{table}
\begin{center}
\begin{tabular}{|c|c|c|c|c|}
\hline
 ~ & ~~~M~~~ & $d_{H-T}$, \AA & ~~~~$q$, $e^{-}$~~~~ & $\Delta \varepsilon_F$, eV \\
\hline
  \multirow{3}{*}{1H1T-slab} & Pb &   2.96   &    0.076    & 0 \\
\cline{2-5}
                             & Bi &   2.61   &    0.169    & 0.11 \\
\cline{2-5}
                             & La &   2.39   &    0.32     & 0.26 \\
\hline
  \multirow{3}{*}{1H1T-I}   & Pb &   2.855  &    0.174    &    0 \\
\cline{2-5}
                             & Bi &   2.598  &    0.337    &    0.22 \\
\cline{2-5}
                             & La &   2.346  &    0.602    &   1.10 \\
\hline
  \multirow{3}{*}{1H1T-II}  & Pb &   2.864  &    0.171    &    0 \\
\cline{2-5}
                             & Bi &   2.598  &    0.336    &    0.22 \\
\cline{2-5}
                             & La &   2.347  &    0.602    &    1.10 \\
\hline
\end{tabular}
\end{center}
\caption{\label{tab:DFT}DFT-derived parameters for the considered structures with different MSe tetragonal layers: averaged distance between T and H layers ($d_{H-T}$); averaged charge transfer ($q$) from T to H layers per NbSe$_2$ primitive cell; Fermi level shift ($\Delta \varepsilon_F$).}
\end{table}

At a coarse-grained leve, all spectra presented in Fig.~\ref{fig:DFT_1H1T} resemble that of the NbSe$_2$ monolayer shown in Fig.~\ref{fig:spectra_free}(a). It is essential, however, to analyze the structure and origin of the spectral splittings in detail: a comparative analysis of all the spectra presented enables us to extract the structure and parameters of the interlayer coupling. The main features of the spectral splittings observed in Fig.~\ref{fig:DFT_1H1T} can be summarized as follows.

(i)~In 1H1T-I MLCs with M = Bi, Pb, no appreciable spin-orbit splitting is observed at the K point, in stark contrast to the NbSe$_2$ monolayer. For M = La, however, the splitting is present. In all 1H1T-II and 1H1T-slab structures, the K-point splitting is present and does not differ markedly from that of the monolayer. Since the spin-orbit splitting at K is directly responsible for the Ising protection of superconductivity in high magnetic fields, this dependence of the electronic structure on the relative orientation of adjacent trigonal layers---as well as on the presence and chemical composition of the tetragonal layers---calls for a systematic explanation and description.

(ii)~The second important feature of the studied MLCs is the splitting at the $\Gamma$ point, which is present in the 1H1T-slab and 1H1T-II structures but absent in 1H1T-I MLCs. This observation implies that the $\Gamma$-point splitting is a consequence of having two trigonal layers in the unit cell and corresponds to branches with $p_z=0$, $\pi/c$, where $c$ is the distance between the trigonal layers. It should be noted that this splitting is not equally pronounced in all 1H1T-II MLCs. For instance, in the structure with M = Pb [see Fig.~\ref{fig:DFT_1H1T}(i)], the second branch is only weakly visible. This is a consequence of the unfolding procedure, since the spectrum shown corresponds to $p_z=0$. Indeed, if the adjacent (MSe)$_{1.14}$NbSe$_2$ blocks were arranged with strict translational invariance along the out-of-plane direction with period $c$, the branch corresponding to $p_z = \pi/c$ would have zero projection onto the $p_z = 0$ state represented in the spectrum and would therefore be completely invisible. Consequently, the intensity of this branch depends sensitively on the composition of the specific structure under consideration. The importance of the $\Gamma$-point splitting lies in the fact that it provides access to the magnitude of the interlayer coupling and, in some cases, allows one to draw unambiguous conclusions about the presence or absence of a spin-orbit contribution.

Our next objective is to build on the general symmetry considerations presented in Sec.~\ref{sec:symmetry_analysis} and construct a model that captures all these spectral features. First, it is important to note that, owing to the presence of the tetragonal layer (T) between the trigonal layers (H), all MLCs considered in this work lack both inversion symmetry $I$ and mirror symmetry $\sigma_h$. Consequently, these symmetries impose no restrictions on the allowed types of interlayer coupling, in stark contrast to the TMD bilayers discussed in the previous section. 

The 1H1T-slab MLC is therefore described by the Hamiltonian \eqref{H2} with the most general interlayer coupling \eqref{V_tr}. The corresponding dispersion relation takes the form:
\begin{align}
    \varepsilon_{\rm 1H1T-slab}^\sigma=\zeta+\alpha\sigma\pm \sqrt{(v+\nu\sigma)^2+(w+\lambda\sigma)^2} .
    \label{e_slab}
\end{align}
The spectra of the 1H1T-slab MLCs, displayed in the left column of Fig.~\ref{fig:DFT_1H1T}, show that the spin splitting at the K-point coincides with that of the NbSe$_2$ monolayer, $2\alpha_{\rm 1H\mbox{-}NbSe_2}(\bm{p}_{\rm K})$ [see Fig.~\ref{fig:spectra_free}(a)], at least within the accuracy accessible from the unfolded data. Since our model is built on the premise that the intralayer parameters of a 1H1T-MLC should remain close to those of the isolated monolayer---apart from the appropriate shift of the chemical potential due to charge transfer from the tetragonal layer---we set $\alpha_{\rm 1H1T\mbox{-}slab}(\bm{p}) = \alpha_{\rm 1H\mbox{-}NbSe_2}(\bm{p})$ throughout the Brillouin zone. Guided by the same physical reasoning as in the previous section, we further set $w(\bm{p}) = 0$. 

\begin{figure}[!tbh]
\includegraphics[width=0.8\columnwidth]{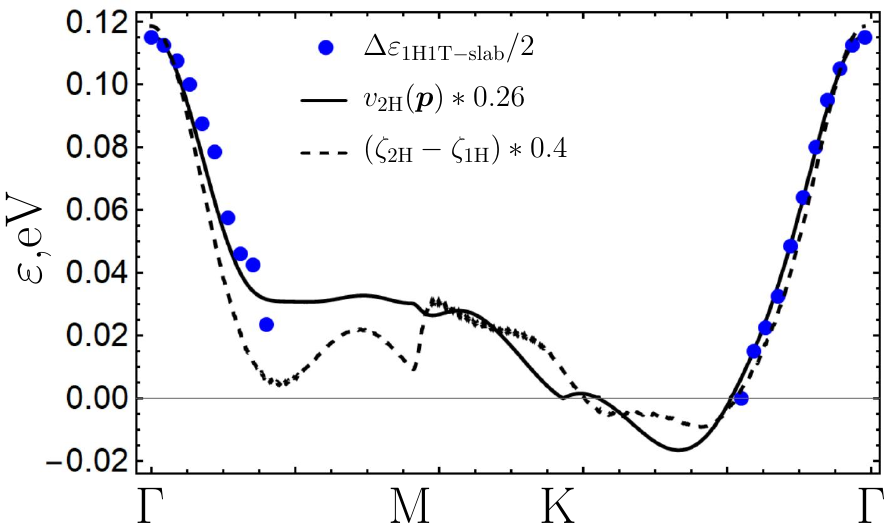}
\caption{Momentum dependence of the energy splitting $\Delta\varepsilon(\bm{p})$ extracted from the 1H1T-slab MLC spectra (blue circles), compared with the scaled interlayer hopping of the 2H-NbSe\(_2\) bilayer (solid black line) and the difference between the spin-independent parts of the 2H and 1H-NbSe\(_2\) dispersions (dashed black line).} 
 \label{fig:compare_splitting}
\end{figure}

In addition to the spin-orbit splitting at the K point---which is accompanied by a small additional splitting due to the two-layer structure of NbSe$_2$---the 1H1T-slab MLC exhibits a large splitting at the $\Gamma$ point. This splitting, 
\begin{align}
\Delta\varepsilon_{\rm 1H1T\mbox{-}slab}(\bm{p}_\Gamma) = 2\sqrt{v(\bm{p}_\Gamma)^2 + \lambda(\bm{p}_\Gamma)^2},
\label{splitting_gamma_slab}
\end{align}
is clearly of spin-independent origin and arises from the presence of two layers, since $\alpha(\bm{p}_\Gamma) = \nu(\bm{p}_\Gamma) = 0$ by virtue of their momentum-odd character. Fig.~\ref{fig:compare_splitting} shows $\Delta\varepsilon_{\rm 1H1T\mbox{-}slab}(\bm{p})$ as a function of momentum along the $\Gamma$--M--K--$\Gamma$ path. Its functional form is in good agreement with that of $v_{\rm 2H}(\bm{p})$, exhibiting a maximum at $\Gamma$ and a minimum in the vicinity of the K point. This common momentum dependence of the even-in-momentum interlayer hopping terms is supported by earlier studies~\cite{Cappelluti2013}, where it was found that the interlayer hopping in multilayer NbSe$_2$ is dominated by the overlap of Se $p_z$~orbitals, which naturally peaks at $\Gamma$. Based on the data presented in Fig.~\ref{fig:compare_splitting}, in the framework of our model we therefore set
\begin{align}
v(\bm{p}) = \gamma_v v_{\rm 2H}(\bm{p}), ~~
\lambda(\bm{p}) = \gamma_\lambda v_{\rm 2H}(\bm{p}),
\label{gamma_even_def}
\end{align}
where $\gamma_v$ and $\gamma_\lambda$ are momentum-independent coefficients. The relative magnitude of $\gamma_v$ and $\gamma_\lambda$ cannot be determined solely from the available DFT data for the 1H1T-slab MLCs. To resolve this, we have also performed DFT calculations with spin-orbit coupling switched off. For M = Bi, the splitting at the $\Gamma$ point with SOC is $\Delta\varepsilon_{\rm 1H1T\mbox{-}slab}^{\rm SOC}(\bm{p}_\Gamma) = 0.38$~eV, while without SOC it is $\Delta\varepsilon_{\rm 1H1T\mbox{-}slab}^{\rm noSOC}(\bm{p}_\Gamma) = 0.33$~eV. If this difference is attributed to the interlayer spin-orbit coupling $\lambda(\bm p)$ rather than to a renormalization of the ordinary interlayer coupling $v(\bm{p})$, Eq.~\eqref{splitting_gamma_slab} yields the estimate
\begin{align}
&\frac{\lambda(\bm{p}_\Gamma)}{v(\bm{p}_\Gamma)} = \frac{\gamma_\lambda}{\gamma_v} \nonumber \\
&= \sqrt{\frac{[\Delta\varepsilon_{\rm 1H1T\mbox{-}slab}^{\rm SOC}(\bm{p}_\Gamma)]^2-[\Delta\varepsilon_{\rm 1H1T\mbox{-}slab}^{\rm noSOC}(\bm{p}_\Gamma)]^2}{[\Delta\varepsilon_{\rm 1H1T\mbox{-}slab}^{\rm SOC}(\bm{p}_\Gamma)]^2}} \approx 0.6 .
\label{lambda_slab}
\end{align}
For significantly smaller differences between the splittings with and without SOC, the accuracy attainable with the unfolding procedure does not permit a reliable estimate of $\lambda$ by this method. For example, $\lambda=0.3v$ yields a difference of only 4\% between $\Delta\varepsilon_{\rm 1H1T\mbox{-}slab}^{\rm SOC}(\bm{p}_\Gamma)$ and $\Delta\varepsilon_{\rm 1H1T\mbox{-}slab}^{\rm noSOC}(\bm{p}_\Gamma)$. The DFT spectra of the 1H1T-slab MLCs with M = Pb and La show no appreciable difference between the results with and without SOC. This suggests that $\lambda$, if present in these compounds, is smaller in magnitude than for M = Bi.

We note in passing that, as seen from the data in Fig.~\ref{fig:spectra_free}, the spin-independent parts of the spectra of 1H-NbSe$_2$ and 2H-NbSe$_2$, denoted $\zeta_{\rm 1H}$ and $\zeta_{\rm 2H}$ respectively, are slightly different. While this difference can be neglected near the K point, it is more pronounced around $\Gamma$. The difference $\zeta_{\rm 2H} - \zeta_{\rm 1H}$ is also plotted in Fig.~\ref{fig:compare_splitting} as a function of momentum; it closely matches the functional form of $v_{\rm 2H}(\bm{p})$. For the moment we merely note this observation; its physical content will be discussed later, in Sec.~\ref{sec:tight-binding}. 

The Schr\"{o}dinger equation for the two-layer wavefunction $(\Psi_{1}^\sigma(\bm{p}), \Psi_{2}^\sigma(\bm{p}))^T$ with the Hamiltonian \eqref{H2} is naturally generalized to bulk MLCs. Specifically for 1H1T-I MLCs the generalized equation reads as follows:
\begin{align}
   (\zeta+\sigma\alpha)\Psi_n^\sigma+V^\sigma \Psi_{n+1}^\sigma+ V^{\sigma *} \Psi_{n-1}^\sigma=\varepsilon^\sigma \Psi_n^\sigma,
    \label{Sr_1H1T_simp}
\end{align}
where $\Psi_n^\sigma(\bm p)$ is a wave function for the $n$-th NbSe$_2$ layer. Applying periodic boundary conditions $\Psi_n^\sigma = \Psi_{n+1}^\sigma$ (which corresponds to setting the out-of-plane momentum $p_z = 0$) to Eq.~\ref{Sr_1H1T_simp} with $\hat{V}^\sigma$ given by Eq.~\eqref{V_tr}, one obtains:
\begin{align}
    \varepsilon_{\rm 1H1T\mbox{-}I}^\sigma =
        \zeta + \alpha\sigma + 2(v + \nu\sigma).
    \label{e_1H1T_simp}
\end{align}
Here, the terms proportional to $\lambda$ and $w$ cancel, while those proportional to $v$ and $\nu$ double. The spectra presented in the middle column of Fig.~\ref{fig:DFT_1H1T} show that, in stark contrast to the NbSe$_2$ monolayer, the Bi- and Pb-based 1H1T-I MLCs exhibit no appreciable spin-orbit splitting anywhere in the Brillouin zone. To satisfy this condition, we must impose $\nu(\bm{p}) \approx -\alpha(\bm{p})/2$. We are thus led to the conclusion that in bulk MLCs, the presence of two adjacent tetragonal layers surrounding each trigonal layer generates an interlayer ISOC. At the same time, the La-based 1H1T-I MLC does exhibit a spin-orbit splitting at the K point [see Fig.~\ref{fig:DFT_1H1T}(b)], indicating that the interlayer spin-orbit coupling induced by the LaSe layer is weak compared to that induced by the Bi- and Pb-based tetragonal layers. This is to be expected, since La is a lighter element with substantially weaker atomic spin-orbit coupling.

In the 1H1T-II MLC, the unit cell hosts two NbSe$_2$ layers. The system is therefore described by a two-component wavefunction $(\Psi_{n,A}^\sigma(\bm{p}), \Psi_{n,B}^\sigma(\bm{p}))^T$, whose Schr\"{o}dinger equation reads:
\begin{align}
   (\zeta+\sigma\alpha)\Psi_{n,A}^\sigma+V^\sigma \Psi_{n,B}^\sigma+ V^{\sigma *} \Psi_{n-1,B}^\sigma=\varepsilon^\sigma \Psi_{n,A}^\sigma, \nonumber \\
   (\zeta-\sigma\alpha)\Psi_{n,B}^\sigma+V^\sigma \Psi_{n+1,A}^\sigma+ V^{\sigma *} \Psi_{n,A}^\sigma=\varepsilon^\sigma \Psi_{n,B}^\sigma .
    \label{Sr_1H1T_true}
\end{align}
The electronic dispersion law at $p_z=0$ takes the form:
\begin{align}
    \varepsilon_{\rm 1H1T-II}^\sigma=\zeta\pm\sqrt{\alpha^2+4(v+\nu\sigma)^2} .
    \label{e_1H1T_true}
\end{align}
The splitting at the $\Gamma$ point takes the form $\Delta\varepsilon(\bm{p}_\Gamma) = 4v(\bm{p}_\Gamma)$, which allows us to determine $\gamma_v$. For the MLCs shown in Fig.\ref{fig:DFT_1H1T}, one can estimate $\gamma_v({\rm La})=-0.10$, $\gamma_v({\rm Pb})=0.24$, $\gamma_v({\rm Bi})=0.42$.

\begin{figure}[!tbh]
\includegraphics[width=\columnwidth]{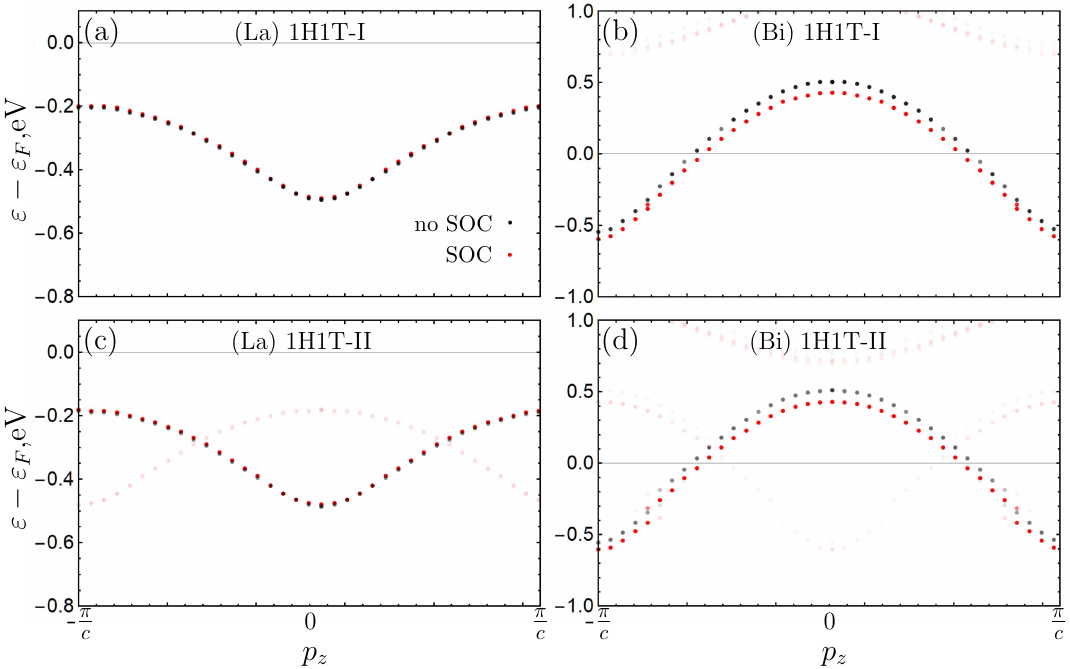}
\caption{Out-of-plane electronic band dispersions of the MLCs 1H1T‑I (M = La, Bi) in (a) and (b), and 1H1T‑II (M = La, Bi) in (c) and (d). The red and black curves correspond to calculations with and without spin–orbit coupling, respectively.} 
 \label{fig:outofplane}
\end{figure}

As can be seen from Eqs.~\eqref{e_1H1T_simp} and \eqref{e_1H1T_true}, the in-plane energy spectrum at $p_z = 0$ for bulk 1H1T MLCs is independent of parameter $\lambda$. This parameter, however, can be extracted from the out-of-plane band dispersion. Within the framework of the present model, the dispersion along the out-of-plane direction at the $\Gamma$ point takes the
form
\begin{align}
    \varepsilon_{\rm 1H1T-I}^\sigma(p_z) =
        \zeta(0) + 2\bigl(v\cos(p_z c) - \sigma\lambda\sin(p_z c)\bigr), \nonumber\\ 
    \varepsilon_{\rm 1H1T-II}^\sigma(p_z) =
        \zeta(0) \pm 2\bigl(v\cos(p_z c) - \sigma\lambda\sin(p_z c)\bigr).
    \label{1H1T_II_z}
\end{align}
As can be seen from Eq.~\eqref{1H1T_II_z}, if $\lambda = 0$ the dispersion $\varepsilon^\sigma(p_z)$ is an even function of $p_z$. For $\lambda \neq 0$ there is a spin splitting of the spectrum branches along $p_z$ and the symmetry $\varepsilon^\sigma(p_z) = \varepsilon^\sigma(-p_z)$ is lost. Fig.~\ref{fig:outofplane} shows the electronic spectra along the $p_z$-direction for the 1H1T-I and 1H1T-II MLCs with M = La, Bi. The data indicate the absence of $\lambda$ in these compounds. Accordingly, we set $\gamma_\lambda = 0$ for all bulk MLCs in what follows. Fig.~\ref{fig:outofplane} further demonstrates that the dispersion along the $p_z$ axis is electron-like for (LaSe)$_{1.14}$NbSe$_2$, whereas for (BiSe)$_{1.14}$NbSe$_2$ it exhibits hole-like character. Consequently, as already indicated above, this yields $\gamma_v<0$ for (LaSe)$_{1.14}$NbSe$_2$ and $\gamma_v>0$ for (BiSe)$_{1.14}$NbSe$_2$.

In summary, within the framework of our model, the interlayer coupling takes the form
\begin{align}
    \hat{V}^\sigma &= v(\bm p)- i\sigma\lambda(\bm{p})+\sigma\nu(\bm{p}) 
    \nonumber \\
    &= (\gamma_v - i\gamma_{\lambda}\,\sigma)  v_{\rm 2H}(\bm{p}) + \gamma_{\nu}\,\sigma\,
       \alpha_{\rm 1H\mbox{-}NbSe_2}(\bm{p}).
          \label{interlayer_1H1T}
\end{align}
The second equality reflects the fact that the momentum dependence of the even-in-momentum interlayer coupling terms $v(\bm{p})$ and $\lambda(\bm{p})$ follows that of $v_{\rm 2H}(\bm{p})$, while the momentum dependence of $\nu(\bm{p})$---odd in momentum---follows that of $\alpha_{\rm 1H\mbox{-}NbSe_2}(\bm{p})$. This allows us to introduce the momentum-independent coefficients $\gamma_v$, $\gamma_\lambda$, and $\gamma_\nu$, which depend on the composition of the tetragonal layer and, generally, on the specific type of structure (slab vs.\ bulk), since the spacing between adjacent tetragonal and trigonal layers  differs in the two cases. This choice of model parameters successfully describes the spectra of all 1H1T MLCs studied in this work. Specific examples of fits to the DFT spectra are presented in Sec.~\ref{sec:tight-binding}.

It should be noted that $v_{\rm 2H}(\bm p)$ in the interlayer coupling expression \eqref{interlayer_1H1T} is used for MLCs of all types, irrespective of whether the adjacent NbSe$_2$ layers separated by a tetragonal layer exhibit a 2H-like or a 1H$+$1H-like mutual stacking. In real MLCs, the 1H$+$1H stacking of trigonal layers occurs precisely in this situation---i.e., across a tetragonal layer---and is not realized within a single trigonal block. The 1H$+$1H bilayer represents a somewhat special case, likely because the mean interlayer spacing in this stacking configuration is significantly smaller than in 2H-NbSe$_2$. As a result, the interlayer coupling $v_{\rm 1H+1H}(\bm{p})$ is systematically larger than $v_{\rm 2H}(\bm{p})$ throughout the Brillouin zone---to a rough approximation, by a constant offset; see Fig.~\ref{fig:spectra_free}. For MLCs, where such small interlayer spacings between trigonal layers are not realized, $v_{\rm 2H}(\bm p)$ is therefore the more appropriate choice.

\section{DFT results and effective model for 2H1T $\mathrm{\textbf{MLCs}}$}
\label{sec:1T2H}

\begin{figure}[!tbh]
\includegraphics[width=\columnwidth]{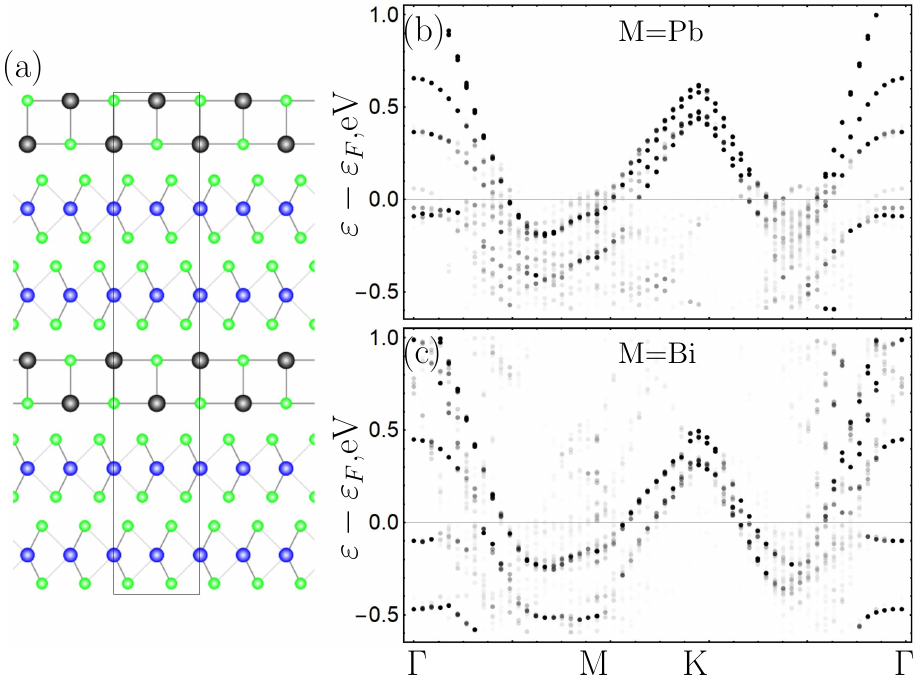}
\caption{(a) Atomic structure of the 2H1T misfit layered compound (side view), consisting of two trigonal NbSe\(_2\) layers interleaved with a single tetragonal MSe layer. (b) and (c) Unfolded DFT-calculated electronic band structures near the Fermi level for the 2H1T MLCs with M = Pb and M = Bi, respectively. The spectra are plotted along the high-symmetry path \(\Gamma\)-M-K-\(\Gamma\).} 
 \label{fig:1T2H_DFT}
\end{figure}

In this section, we turn to more complex bulk MLCs in which each trigonal block consists of two NbSe$_2$ layers; see Fig.~\ref{fig:1T2H_DFT}(a). The DFT-calculated electronic spectra of 2H1T-(PbSe)$_{1.14}$(NbSe$_2$)$_2$ and 2H1T-(BiSe)$_{1.14}$(NbSe$_2$)$_2$ are presented in Figs.~\ref{fig:1T2H_DFT}(b) and (c), respectively. DFT results for some other 2H1T MLCs, both bulk and surface, are available in the literature, see for example ~\cite{Ng2022,Leriche2021,Samuely2023,Zullo2023}. A qualitative inspection of the spectra in Figs.~\ref{fig:1T2H_DFT}(b) and (c) reveals that the electronic structure of these compounds exhibits a strong splitting in the vicinity of the K-point, and is thus closer to that of the 1H-NbSe$_2$ monolayer than to that of the 2H-NbSe$_2$ bilayer [cf.\ Figs.~\ref{fig:spectra_free}(a) and (b)]. At the same time, and in contrast to 1H-NbSe$_2$, a four-fold splitting of the spectral branches is observed near the $\Gamma$-point.

Our aim here is to demonstrate, using 2H1T MLCs as an example, how the model constructed in the previous section for 1H1T MLCs is generalized to more complex compounds. In the next section, we will develop a fully quantitative description of all model parameters within the tight-binding approximation, assess the accuracy with which the model reproduces our DFT spectra, and present model-based spectral calculations for several MLCs.

In the 2H1T MLC, the unit cell hosts four NbSe$_2$ layers, see Fig.~\ref{fig:1T2H_DFT}(a). The system is therefore described by a four-component wavefunction $(\Psi_{n,A}^\sigma(\bm{p}), \Psi_{n,B}^\sigma(\bm{p}), \Psi_{n,C}^\sigma(\bm{p}), \Psi_{n,D}^\sigma(\bm{p}))^T$, whose Schr\"{o}dinger equation reads:
\begin{align}
    (\zeta+\alpha\sigma)\Psi^\sigma_{n,A}+v_{\rm 2H}\Psi^\sigma_{n,B}+V^{\sigma*}\Psi^\sigma_{n-1,D}&=\varepsilon^\sigma \Psi^\sigma_{n,A} \nonumber \\
    (\zeta-\alpha\sigma)\Psi^\sigma_{n,B}+v_{\rm 2H}\Psi^\sigma_{n,A}+V^{\sigma*}\Psi^\sigma_{n,C}&=\varepsilon^\sigma \Psi^\sigma_{n,B} \\ \nonumber
    (\zeta-\alpha\sigma)\Psi^\sigma_{n,C}+v_{\rm 2H}\Psi^\sigma_{n,D}+V^{\sigma}\Psi^\sigma_{n,D}&=\varepsilon^\sigma \Psi^\sigma_{n,C} \\ \nonumber
    (\zeta+\alpha\sigma)\Psi^\sigma_{n,D}+v_{\rm 2H}\Psi^\sigma_{n,C}+V^{\sigma}\Psi^\sigma_{n+1,A}&=\varepsilon^\sigma \Psi^\sigma_{n,D} ,
\end{align}
where $V^\sigma$ is described by Eq.~(\ref{interlayer_1H1T}). 

\section{Tight-binding representation of the MLC effective model}
\label{sec:tight-binding}

\begin{figure}[!tbh]
\includegraphics[width=0.7\columnwidth]{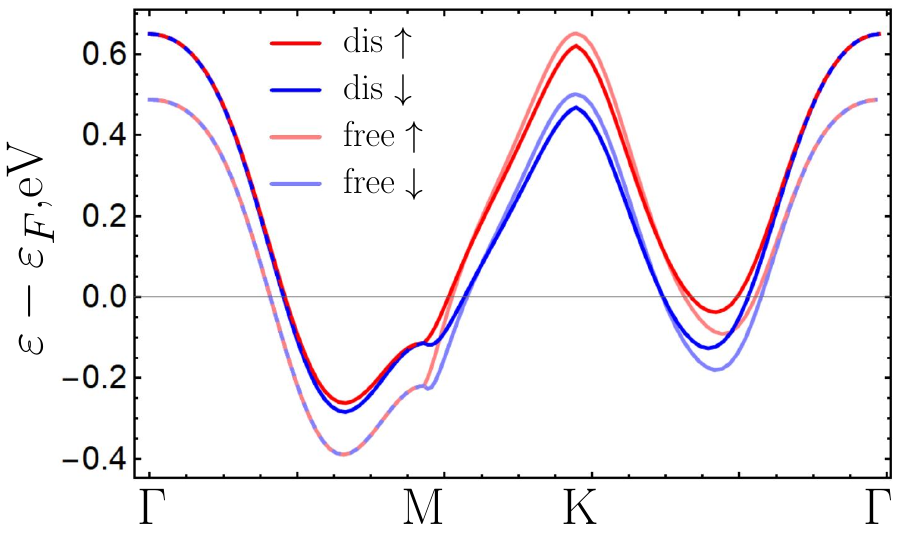}
\caption{Comparison of the DFT-calculated electronic dispersions along the high-symmetry path $\Gamma$--M--K--$\Gamma$ for a free-standing 1H-NbSe$_2$ monolayer, $\varepsilon_{\rm 1H}^\sigma(\bm{p}) = \zeta_{\rm 1H}(\bm{p}) + \sigma\alpha(\bm{p})$ (labeled ``free''), and for a NbSe$_2$ monolayer subjected to the uniaxial tensile strain characteristic of its environment within the misfit structure,
$\varepsilon_{\rm dis}^\sigma(\bm{p}) =
\zeta_{\rm dis}(\bm{p}) + \sigma\alpha(\bm{p})$
(labeled ``dis''). } 
 \label{fig:1H_distorted}
\end{figure}

The intralayer dispersion of a 1H-NbSe$_2$ layer is described by the tight-binding Hamiltonian
\begin{align}
    \hat{H}_{\rm kin} = \sum_{\bm{i}\bm{j},\sigma}
        t_{\bm{i}\bm{j},\sigma} c_{\bm{i}\sigma}^{\dagger} c_{\bm{j}\sigma},
    \qquad t_{\bm{i}\bm{j},\sigma} = t_{\bm{j}\bm{i},\sigma}^*,
    \label{tb_intralayer}
\end{align}
where $c_{\bm{i}\sigma}$ annihilates an electron at Nb site $\bm{i}$ with spin $\sigma$, and $t_{\bm{i}\bm{j},\sigma}$ is the hopping amplitude between sites $\bm{i}$ and $\bm{j}$. The hopping matrix element is decomposed into spin-independent and spin-orbit parts as $t_{\bm{i}\bm{j},\sigma} = t_{\bm{i}\bm{j}}^\zeta + i\sigma t_{\bm{i}\bm{j}}^\alpha$, which in momentum space yield
\begin{align}
    \zeta(\bm{p}) &= \sum_{\bm{i}\bm{j}}
        t_{\bm{i}\bm{j}}^\zeta e^{i\bm{p}\cdot(\bm{i}-\bm{j})}, \\
    \alpha(\bm{p}) &= \sum_{\bm{i}\bm{j}}
        t_{\bm{i}\bm{j}}^\alpha e^{i\bm{p}\cdot(\bm{i}-\bm{j})}
        e_{\bm{i}\bm{j}},
\end{align}
where $e_{\bm{i}\bm{j}} = \pm 1$ is antisymmetric under a $60^\circ$ rotation and symmetric under a $120^\circ$ rotation, reflecting the $C_3$ rotational symmetry and broken six-fold symmetry of the NbSe$_2$ monolayer. Following previous studies~\cite{Aikebaier2022,Bobkov2024_proximity} on the NbSe$_2$ monolayer, we adopt a model that includes six nearest-neighbor spin-independent hopping parameters $t_{1-6}^\zeta$, an on-site energy $t_0^\zeta$, and the spin-orbit hopping terms $t_{1,3}^\alpha$. The spin-orbit terms $t_{0,2,6}^\alpha$ are forbidden by symmetry. The remaining terms $t_{4,5}^\alpha$ are omitted because the set of hopping parameters already included is sufficient to achieve excellent quantitative agreement between the model and the DFT dispersion of the NbSe$_2$ monolayer.

\begin{figure*}[!tbh]
\includegraphics[width=1.8\columnwidth]{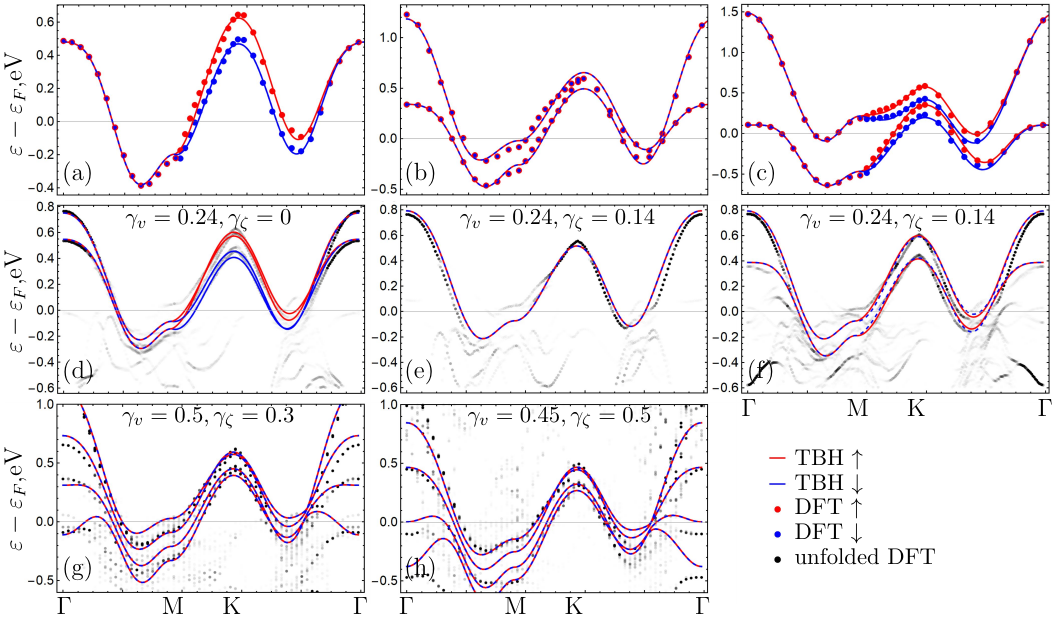}
\caption{Comparison of the DFT-calculated electronic band structures (points) and the corresponding tight-binding model (solid lines) for various NbSe$_2$-based misfit layered compounds along the high-symmetry path $\Gamma$--M--K--$\Gamma$. Red and blue denote spin-up ($\uparrow$) and spin-down ($\downarrow$) states, respectively. The panels correspond to the following systems: (a)~1H-NbSe$_2$ monolayer; (b)~2H-NbSe$_2$ bilayer; (c)~1H$+$1H bilayer; (d)~1H1T-slab (Pb); (e)~1H1T-I bulk  (Pb); (f)~1H1T-II bulk  (Pb); (g)~2H1T-(PbSe)$_{1.14}$(NbSe$_2$)$_2$; and (h)~2H1T-(BiSe)$_{1.14}$(NbSe$_2$)$_2$. The unfolded DFT bands are shown as black dots in panels (d)--(h). For all structures (d-h) $\gamma_\lambda=0$, $\gamma_\nu=-0.5$. The specific parameters $\gamma_v$ and $\gamma_\zeta$ are indicated in the figures.} 
 \label{fig:DFT_TBH}
\end{figure*}

As noted above, in MLCs the NbSe$_2$ monolayer is slightly stretched along one of its in-plane axes to match the lattice constant of the adjacent tetragonal layer. This distortion alone modifies the electronic dispersion, even before the full misfit structure is taken into account. Fig.~\ref{fig:1H_distorted} compares the dispersion of a free-standing NbSe$_2$ monolayer, $\varepsilon_{\rm 1H}^\sigma(\bm{p}) = \zeta_{\rm 1H}(\bm{p}) + \sigma\alpha(\bm{p})$, with that of a monolayer subjected to the same uniaxial tensile strain as in the misfit configuration, $\varepsilon_{\rm dis}^\sigma(\bm{p}) = \zeta_{\rm dis}(\bm{p}) + \sigma\alpha(\bm{p})$. Accordingly, the hopping parameters used for $\zeta(\bm{p})$ in 1H1T MLCs, denoted $t_{0-6,{\rm dis}}^{\zeta}$, differ from those of the free-standing monolayer. Moreover, the data presented in Fig.~\ref{fig:1H_distorted} show that the difference $\zeta_{\rm 1H}(\bm{p}) - \zeta_{\rm dis}(\bm{p})$ follows the same function $v_{\rm 2H}(\bm{p})$ as the difference $\zeta_{\rm 1H}(\bm{p}) - \zeta_{\rm 2H}(\bm{p})$, as discussed in Sec.~\ref{sec:1T1H}. We therefore adopt the more general form $\zeta(\bm{p}) = \zeta_{\rm dis}(\bm{p}) - \gamma_\zeta v_{\rm 2H}(\bm{p})$ for the intralayer dispersion, where $\gamma_\zeta$---together with the previously introduced $\gamma_v$, $\gamma_\nu$, and $\gamma_\lambda$---accounts for the renormalization of the intralayer dispersion in the misfit configuration.  

The spin-orbit hopping parameters $t_{1,3}^\alpha$, however, are taken to be identical for the free-standing and distorted monolayers, as this choice already yields excellent quantitative agreement with the DFT spectra. The hopping parameters for the free-standing monolayer $t_{0-6}^\zeta$, the distorted monolayer $t_{0-6, \rm {dis}}^{\zeta}$, and the spin-orbit terms $t_{1,3}^\alpha$ are listed in Tab.~\ref{tab:hopping_par}. These parameters were obtained by minimizing the root-mean-square deviation between the electronic dispersion of the model and the DFT-calculation. 

\begin{table}
\begin{center}
\begin{tabular}{|c|c|c|c|c|c|c|c|c|c|}
\hline
\# & $0$ & $1$ & $2$ & $3$ & $4$ & $5$ & $6$ \\
\hline
$t^\zeta_{\rm 1H}$ & 10.2 & 17.3 & 99.0 & -7.0 & -8.8 & -13.9 & 0.8 \\
\hline
$t^\zeta_{\rm 2H}$ & 47.6 & 26.5 & 109.0 & 7.3 & -6.4 & -10.0 & -0.7 \\
\hline
$t^\zeta_{\rm 1H+1H}$ & -8.8 & 55.6 & 93.0 & 13.9 & -12.1 & -9.6 & 5.0 \\
\hline
$t^\zeta_{\rm 1H-dis}$ & 70.8 & 35.4 & 88.0 & -0.5 & -9.7 & -8.6 & 1.6 \\
\hline
$t^\alpha$ & - & 16.7 & - & 1.6 & n.u. & n.u. & - \\
\hline
$t^v_{\rm 2H}$ & 78 & 21.2 & 8.5 & 25.2 & -0.8 & 9.7 & -5.7 \\
\hline
$t^v_{\rm 1H+1H}$ & 253.8 & 41.1 & 0.2 & 25.9 & -1.4 & 5.0 & 3.1 \\
\hline
\end{tabular}
\end{center}
\caption{\label{tab:hopping_par}Parameters of the one-band tight-binding model. All values are given in meV.}
\end{table} 

The model constructed above is now generalized to structures with multiple trigonal layers. In the tight-binding representation, the coupling between two neighboring trigonal layers labeled by $m$ and $n$ takes the form
\begin{align}
    \hat{V}_{mn} = \sum_{\bm{i}\bm{j},\sigma}
        t_{\bm{i}m,\bm{j}n,\sigma}^v c_{\bm{i}m\sigma}^{\dagger}
        c_{\bm{j}n\sigma},
    \qquad t_{\bm{i}m,\bm{j}n,\sigma}^v = t_{\bm{j}n,\bm{i}m,\sigma}^{v*},
    \label{tb_interlayer}
\end{align}
where the interlayer hopping matrix element is decomposed into spin-independent and spin-orbit contributions as $t_{\bm{i}m,\bm{j}n,\sigma}^v = t_{\bm{i}m,\bm{j}n}^v + \sigma\, t_{\bm{i}m,\bm{j}n}^\nu + \sigma\, t_{\bm{i}m,\bm{j}n}^\lambda$.

For the 2H-NbSe$_2$ bilayer, only the spin-independent hopping $t_{{\rm 2H},0-6}^v$ is non-zero, in accordance with the results of Fig.~\ref{fig:spectra_free}. The corresponding values, obtained by fitting $v_{\rm 2H}(\bm{p})$, are listed in Table~\ref{tab:hopping_par}. For the 1H$+$1H-NbSe$_2$ bilayer, both $t_{{\rm 1H+1H},0-6}^v$ and the spin-orbit term $t_{{\rm 1H+1H},0-6}^\nu$ are non-zero. The latter, however, is negligibly small and is therefore omitted from the model. The values of $t_{{\rm 1H+1H},0-6}^v$, obtained by fitting $v_{\rm 1H+1H}(\bm{p})$ from Fig.~\ref{fig:spectra_free}, are also listed in Table~\ref{tab:hopping_par}.

For hopping across a tetragonal layer, the spin-independent and spin-orbit hopping elements are expressed, in accordance with Eq.~\eqref{interlayer_1H1T}, through the spin-independent interlayer hopping of the 2H-NbSe$_2$ bilayer and intralayer spin-orbit coupling, respectively:
\begin{align}
    t_{\bm i m \bm j n}^v = \gamma_v t_{{\rm{2H}},\bm i \bm j }^v , ~~  t_{\bm i m \bm j n}^\lambda = \gamma_\lambda t_{{\rm{2H}},\bm i \bm j }^v , ~~
    t_{\bm i m \bm j n}^\nu = \gamma_\nu t_{ \bm i \bm j }^\alpha .
       \label{tb_interlayer_MLC}
\end{align}

Fig.~\ref{fig:DFT_TBH} shows the band structure of the monolayer obtained from DFT (points) together with the tight-binding fit (solid line). The tight-binding model constructed in this work is in excellent qualitative and good quantitative agreement with the DFT results. The only parameters adjusted individually for each structure are $\gamma_\zeta$, $\gamma_v$, $\gamma_\lambda$ and $\gamma_\nu$.

Although the electronic spectrum of 2H1T MLCs in the vicinity of the K point may appear close to that of the NbSe$_2$ monolayer, our analysis reveals that the layers are, in fact, strongly coupled. The NbSe$_2$ layers interact strongly both within each trigonal block and with neighboring trigonal blocks across the intervening tetragonal layer. This leads to a large splitting near the $\Gamma$ point and, more importantly, to the emergence of an interlayer spin-orbit coupling. Within our model, the latter provides a natural explanation for the experimentally observed Ising protection of superconductivity; see Sec.~\ref{sec:ising_protection}.

\section{Ising protection of superconductivity on $\mathrm{\textbf{NbSe}_2}$-based $\mathrm{\textbf{MLCs}}$}
\label{sec:ising_protection}
We now turn to the temperature dependence of the critical field in the superconducting state of the MLCs studied above. Treating superconductivity at the mean-field level, the Hamiltonian of MLC in a magnetic field applied parallel to the layers can be written as
\begin{align}
    \hat{H} &= \sum_{n} \hat{H}_{{\rm kin},n}
              + \sum_{m,n,\sigma} \hat{V}_{mn} \nonumber \\
            &+ \mu_B \sum_{\bm{i},n,\sigma,\sigma'}
              c_{\bm{i}n\sigma}^{\dagger}
              (\bm{H}\cdot\bm{\sigma})_{\sigma\sigma'}
              c_{\bm{i}n\sigma'} \nonumber \\
            &+ \bigl[ \sum_{\bm{i},n} \Delta_{\bm i n}
              c_{\bm{i}n\uparrow}^{\dagger}
              c_{\bm{i}n\downarrow}^{\dagger}
              + \text{H.c.} \bigr],
    \label{ham_SC}
\end{align}
where $\hat{H}_{{\rm kin},n}$ is the intralayer kinetic energy term, described by Eq.~(\ref{tb_intralayer}), and $\hat{V}_{mn}$ is the interlayer interaction described by Eq.~(\ref{tb_interlayer}). The second line represents the Zeeman interaction between electrons and the external magnetic field $\bm H$, which is assumed to be in-plane. $\mu_B$ is the Bohr magneton. In full analogy with the case of layered TMD materials~\cite{Engstrom2026}, the orbital contribution of the magnetic field to the suppression of superconductivity is negligible compared to the Zeeman term. We therefore neglect it in what follows. The third line accounts for superconductivity, which is assumed to be of the spin-singlet $s$-wave type throughout this work. The superconducting order parameter $\Delta_{\bm i n}$ is determined self-consistently via $\Delta_{\bm i n} = \lambda \langle c_{in\downarrow} c_{in\uparrow} \rangle$, where $\lambda$ is the pairing constant. For all bulk structures and symmetric slabs considered here, the order parameter is spatially homogeneous and reduces to a single value $\Delta$. 

The order parameter is obtained by solving the Bogoliubov--de Gennes (BdG) equations numerically. Specifically, we diagonalize the Hamiltonian \eqref{ham_SC} by means of the Bogoliubov transformation:
\begin{align}
c_{\bm i n,\sigma}=\sum\limits_l u_{l,\sigma}^{\bm i n}\hat b_l+v^{\bm i n *}_{l,\sigma}\hat b_l^\dagger . 
\label{bogolubov}
\end{align}
The resulting BdG equations take the form:
\begin{align}
 \sigma \Delta_{\bm i n} v^{ \bm i n}_{l,-\sigma} + \sum\limits_{ \bm i' n'} t_{\bm i n \bm i' n',\sigma} u^{\bm i' n'}_{ l, \sigma} + \mu_B H \sigma u^{\bm i n}_{ l, -\sigma} & = \varepsilon_l u_{l,\sigma}^{\bm i n} \nonumber \\  
\sigma \Delta_{ \bm i n}^* u^{\bm i n}_{l,-\sigma} + \sum\limits_{ \bm i' n'} t_{\bm i n \bm i' n',\sigma}^* v^{\bm i' n'}_{ l, \sigma} + \mu_B H \sigma v^{\bm i n}_{ l, -\sigma} & = -\varepsilon_l v_{l,\sigma}^{\bm i n}. 
\label{bdg}
\end{align}
Using the solutions of the BdG equations, the superconducting order parameter at site $(\bm i,n)$ is given by:
\begin{align}
    \Delta_{\bm i n}=\lambda \sum_{l} \bigl( u_{l,\uparrow}^{ \bm i n}v_{l,\downarrow}^{ \bm i n*}f_l+u_{l,\downarrow}^{ \bm i n}v_{l,\uparrow}^{ \bm i n*}(1-f_l)\bigr),
    \label{self-consistency}
\end{align}
where $f_l=1/(1+{\rm exp}(\varepsilon_l/T))$ is a Fermi–Dirac distribution.

It should be noted that, as a minimal model, we consider a single uniform order parameter, $\Delta$, across the entire Brillouin zone. Several studies have addressed the influence of the specific structure of the pairing interaction, distinguishing between the uniform order parameter adopted here ($\Delta$), an order parameter confined to the valleys around the K-points ($\Delta_{\rm K}$), or two distinct order parameters $\Delta_\Gamma$ and $\Delta_{\rm K}$ on the respective Fermi surfaces around $\Gamma$ and in the K-valleys~\cite{Engstrom2025,Engstrom2026}. While the precise choice of the pairing model is certainly important for clarifying the microscopic mechanism of superconductivity in Ising superconductors, for the question we focus on in the present work---namely, the presence or absence of Ising protection---the specific model is immaterial. We therefore adopt the simplest scenario. The resulting self-consistency equation can be rewritten as 
\begin{align}
    \Delta=\Delta_\Gamma=\Delta_{\rm K}=\lambda F_\Gamma + \lambda F_{\rm K},
\end{align}
which is equivalent to a two-band model with strong interband coupling. Here $F_{\Gamma,{\rm K}}$ is an amplitude of singlet correlations around $\Gamma$ and K points:
\begin{align}
    F_{\Gamma,{\rm K}}=\int\limits_{V_{\Gamma,{\rm K}}} \frac{dp^D}{(2\pi)^D} \sum_{l} \bigl( u_{l,\uparrow}^{n}(\bm p)v_{l,\downarrow}^{n*}(\bm p)f_l \nonumber \\
    +u_{l,\downarrow}^{n}(\bm p)v_{l,\uparrow}^{n*}(\bm p)(1-f_l)\bigr),
\end{align}
where $V_{\Gamma,{\rm K}}$ -- parts of the Brillouin zone corresponding the $\Gamma$ and K points. $D$ is a dimensionality of the system ($=3$ for bulk systems and $=2$ for slab systems); $u_{l,\sigma}^n(\bm p),v_{l,\sigma}^n(\bm p)$ is a Fourier transform of the $u_{l,\sigma}^{\bm i n},v_{l,\sigma}^{\bm i n}$:
\begin{align}
    u(v)_{l,\sigma}^{\bm i n} = \int \frac{dp^D}{(2\pi)^D} u(v)_{l,\sigma}^n (\bm p) e^{i (\bm i +\bm n)\bm p}.
\end{align}
Eq.~\eqref{self-consistency} was solved in the linearized regime with respect to $\Delta$, which corresponds to a second-order phase transition. In purely Pauli-limited superconductors, the presence of a magnetic field can drive the transition first order. However, at finite spin-orbit coupling the spin susceptibility of the superconducting state remains close to its normal-state value, so that the transition remains continuous at all temperatures~\cite{Sohn2018,Frigeri2004,Samokhin2005,Samokhin2007}. It is therefore generally justified to assume a continuous superconducting transition in non-centrosymmetric Ising superconductors. In any case, even if a first-order transition were to occur, neglecting it would, at most, underestimate the critical field at a given temperature~\cite{Sarma1963}, and thus does not affect the conclusion regarding the existence of Ising protection.

\begin{figure}[!tbh]
\includegraphics[width=\columnwidth]{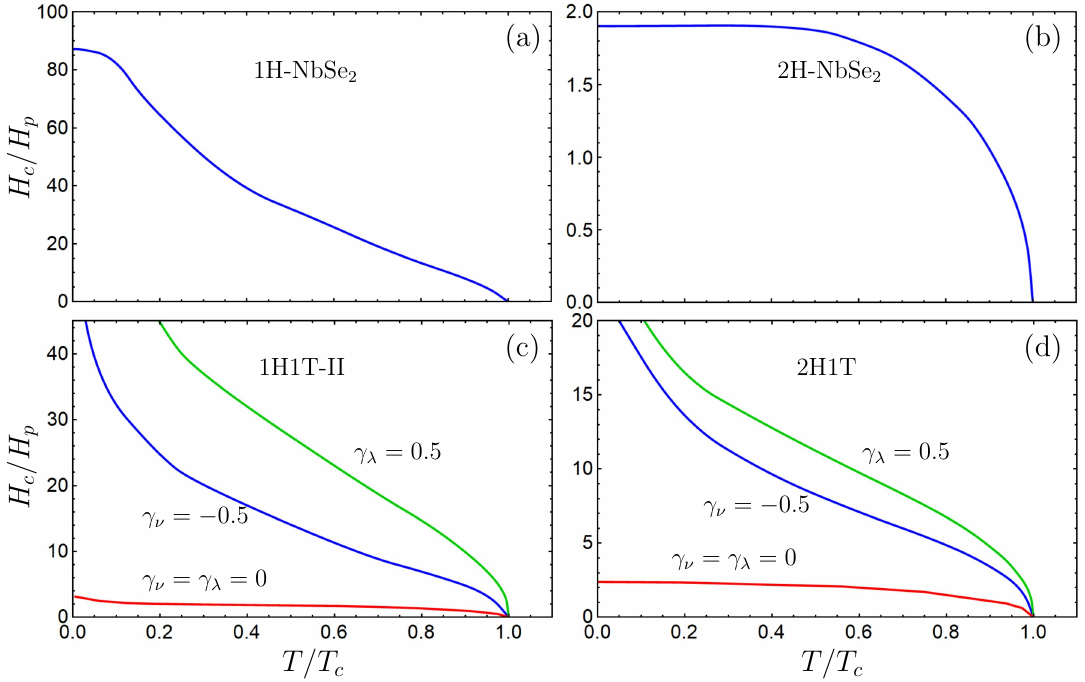}
\caption{Temperature dependence of the normalized in-plane critical field $H_c/H_P$ for various NbSe$_2$-based systems. Here $H_P = \Delta/\sqrt{2}$ is the Pauli paramagnetic limit. (a)~1H-NbSe$_2$ monolayer; (b)~2H-NbSe$_2$ bilayer; (c)~1H1T-II bulk MLC; (d)~2H1T bulk MLC. Panels (c) and (d) illustrate the enhancement of the critical field for different strengths of the interlayer spin-orbit coupling: $\gamma_\nu = -0.5$, $\gamma_\lambda = 0$ (blue curves); $\gamma_\nu = 0$, $\gamma_\lambda = 0.5$ (green curves); and without the interlayer spin-orbit coupling, $\gamma_\nu = \gamma_\lambda = 0$ (red curves). Critical temperature of 3~K was chosen for monolayer systems (a),(c) and 5~K for bilayer systems (b),(d). } 
 \label{fig:Hc}
\end{figure}

The temperature dependence of the critical field $H_c$ at which the superconducting order parameter vanishes is shown in Fig.~\ref{fig:Hc}. The strong enhancement of $H_c$ over the paramagnetic limit $H_P=\Delta/\sqrt{2}$ in 1H-NbSe$_2$ monolayer, shown in Fig.~\ref{fig:Hc}(a), is in qualitative agreement with well-known results for the NbSe$_2$ monolayer \cite{Xi2016,delaBarrera2018,Saito2016,Ilic2017,Mockli2018,Mockli2019,Mockli2020,Engstrom2025,Engstrom2026} (quantitative results are very sensitive to the pairing model, disorder etc.). Owing to the intralayer spin-orbit coupling, which induces a strong spin splitting $\alpha(\bm{p})$ of the normal-state spectrum, Zeeman pair-breaking of singlet Cooper pairs is suppressed up to very high in-plane fields that vastly exceed the Pauli paramagnetic limit $H_P$. Superconductivity is destroyed only when the Zeeman energy $\mu_B H$ becomes comparable to the spin-orbit splitting.

Fig.~\ref{fig:Hc}(b) shows the corresponding results for the 2H-NbSe$_2$ bilayer. Here, the Ising protection of superconductivity is significantly weaker: $H_c(T=0)$ exceeds the Pauli limit by less than a factor of two. This behavior is also well known \cite{delaBarrera2018,Samuely2023,Engstrom2026} and is understood as follows. In the 2H-NbSe$_2$ bilayer, inversion symmetry is restored, and the intralayer spin-orbit term therefore has opposite signs on the two layers [see Eq.~\eqref{ham_2H}]. As a result, the delocalization of the Cooper pair across the two layers weakens the Ising protection. The stronger the interlayer coupling, i.e., the greater the degree of pair delocalization, the weaker the protection. As can be seen from the data presented in Fig.~\ref{fig:spectra_free}, for realistic parameters of the 2H-NbSe$_2$ bilayer the interlayer coupling $v(\bm{p})$ in the vicinity of the K point is of the same order as  the monolayer spin-orbit splitting $\alpha(\bm{p})$. Nevertheless, a partial protection persists even in this regime. It has also been shown that the Ising protection can be enhanced by introducing an antisymmetric on-site energy (chemical potential) shift $\pm\Delta \varepsilon_F$ between the two constituent monolayers~\cite{Samuely2023,Engstrom2026}. For NbSe$_2$ bilayers grown on a substrate, such an asymmetric chemical-potential shift is indeed expected; here, however, we consider a free-standing bilayer.

Fig.~\ref{fig:Hc}(c,d) presents the results for MLCs, comparing $H_c(T)$ obtained with and without the interlayer spin-orbit coupling. In the absence of the interlayer spin-orbit term, our model yields a modest enhancement of $H_c(T)$ over the Pauli limit $H_P$, which is even slightly larger than that found for the 2H-NbSe$_2$ bilayer. This can be understood on physical grounds: in the 1H1T MLC [Fig.~\ref{fig:Hc}(c)], adjacent NbSe$_2$ layers are separated by tetragonal layers, which reduce the interlayer coupling and thereby localize the Cooper pairs more strongly within a single layer. The resulting Ising protection is therefore stronger than in the 2H-NbSe$_2$ bilayer. For the 2H1T MLC [Fig.~\ref{fig:Hc}(d)], the Cooper pair is delocalized not only within a given trigonal block, but also partially extends into the NbSe$_2$ layer of the neighboring trigonal block, which carries the same sign of the intralayer spin-orbit coupling. This further enhances the Ising protection.

The critical-field ratios $H_c(T=0)/H_P$ obtained in this way are considerably smaller than the experimental estimates for (LaSe)$_{1.14}$(NbSe$_2$) and (LaSe)$_{1.14}$(NbSe$_2$)$_2$~\cite{Samuely2021,Samuely2023}. Within our model, this difference can be attributed to the presence of a weak interlayer spin-orbit coupling $\nu$ in these compounds. An alternative hypothesis was put forward in Ref.~\cite{Samuely2023}, suggesting that the experimentally observed enhancement of the critical field in (LaSe)$_{1.14}$(NbSe$_2$)$_2$ could arise from an antisymmetric chemical-potential shift $\pm\Delta \varepsilon_F$ between the two NbSe$_2$ monolayers forming each trigonal block. Such a shift, however, requires a physical origin---for instance, an asymmetric arrangement of the tetragonal layers above and below a given trigonal block. In the compounds studied here, this asymmetry is absent; see Fig.~\ref{fig:1T2H_DFT}(a). Even if MLCs with a different mutual orientation of the tetragonal layers were considered, the resulting antisymmetry would periodically change sign along the incommensurate direction, implying a spatial modulation of $\Delta \varepsilon_F$ along that axis. Accounting for the effect of such a spatially periodic antisymmetric shift lies beyond the scope of the present work.

As can be seen from Figs.~\ref{fig:Hc}(c) and (d), the inclusion of the interlayer spin-orbit coupling can enhance the critical field by a large factor. This enhancement arises both from the term $\nu(\bm{p})$, which is odd in the in-plane momentum, and from $\lambda(\bm{p})$, which is even in the in-plane momentum but odd in the out-of-plane component. For heavy elements such as Pb and Bi, our model yields a substantial odd-in-momentum term $\nu(\bm{p})$ with $\gamma_\nu \approx -0.5$ for bulk MLCs, whereas for La-based tetragonal layers the interlayer spin-orbit coupling is significantly weaker. This leads to a strong enhancement of the Ising protection of superconductivity in Pb- and Bi-based MLCs, while for La-based compounds the interlayer spin-orbit coupling does contribute to an increase of the critical field, albeit to a much lesser extent.

For 1H1T-I MLCs, we have verified that, in the absence of interlayer spin-orbit coupling ($\gamma_\nu = \gamma_\lambda = 0$), the Ising protection of superconductivity remains of the same order as in the monolayer. This is physically expected: in these compounds all trigonal layers share the same orientation, so that delocalization of the Cooper pair between adjacent layers does not reduce the intralayer ISOC. The inclusion of a negative interlayer spin-orbit coupling, $\nu(\bm{p}_{\rm K}) \approx -\alpha(\bm{p}_{\rm K})/2$, as introduced above for MLCs with heavy elements such as Bi and Pb, lowers the ratio $H_c(T=0)/H_P$ by approximately a factor of two, yet the critical field remains high. Although the electronic dispersion of Bi- and Pb-based 1H1T-I MLCs becomes nearly spin-degenerate near the K point [see Figs.~\ref{fig:DFT_1H1T}(e) and (h)], this happens only at $p_z = 0$; the spin splitting persists at all other out-of-plane momenta around K-point, which explains the observed Ising protection.

The analysis above was restricted to the simplest, spin-singlet pairing scenario. For NbSe$_2$ and other TMD Ising superconductors, the admixture of triplet pairing has also been discussed~\cite{Mockli2020,He2018,Wickramaratne2020,Engstrom2026}. In particular, it was shown that, for a fixed magnitude of the singlet order parameter, a triplet admixture enhances the Ising protection of superconductivity. Since the mechanism of Ising protection proposed in the present work for MLCs---interlayer spin-orbit coupling---is fundamentally different from the dominant mechanism in TMD monolayers and few-layer structures, the effect of triplet pairing on the degree of Ising protection in MLCs requires a separate investigation and will be addressed in a future publication.

\section{Conclusions}
\label{sec:conclusions}

In summary, we have developed a unified tight-binding model for the electronic structure of misfit layered compounds (MLCs) based on transition-metal dichalcogenides, focusing on the NbSe$_2$ family. Our approach bridges the gap between the well-understood electronic properties of isolated TMD monolayers and the complex bulk physics of MLCs. The central  result of this work is that the conventional ``rigid-band'' picture---which treats the tetragonal layers merely as charge reservoirs that electronically isolate the trigonal slabs---is qualitatively incomplete.

By systematically analyzing DFT data for several nonmagnetic MLCs, we have demonstrated that the tetragonal layers play a far more active role. They not only mediate the interlayer coupling between trigonal blocks but, crucially, generate an appreciable interlayer spin-orbit coupling. This key ingredient, entirely absent in the standard rigid-band description, is essential for capturing the electronic structure of bulk MLCs. The model, validated against DFT for both 1H1T and 2H1T compounds, naturally accounts for the emergence or suppression of an effective spin-orbit splitting in structures where it is not expected from the isolated building blocks alone.

By solving the Bogoliubov--de Gennes equations, we have further shown that the inclusion of the derived interlayer spin-orbit coupling strongly enhances the in-plane  critical field $H_{c}$, providing a natural mechanism for the Ising protection of superconductivity observed experimentally in these bulk materials. The interlayer spin-orbit coupling acts in addition to the intralayer ISOC, and its inclusion brings the calculated critical fields into agreement with the enhanced values reported experimentally---an agreement that cannot be achieved by models based on electronically isolated layers.

Our work establishes a new theoretical framework for understanding MLCs, showing that they are not merely a mechanical assembly of 2D building blocks but a distinct class of three-dimensional materials with intrinsically coupled layers and emergent spin-orbit phenomena. This framework provides a solid foundation for future studies of the rich variety of correlated states predicted in these systems---including topological superconductivity and layer-selective Fulde-Ferrell-Larkin-Ovchinnikov phases---and offers a clear strategy for engineering their electronic properties through chemical composition and stacking sequence.

\begin{acknowledgments}
We thank Dr. A.S. Frolov and Dr. S.V. Eremeev for stimulating discussions. The work was supported by the Russian Science Foundation via the project No. 25-22-00579. 
\end{acknowledgments}

\bibliography{misfits}

\end{document}